\documentclass[fleqn,usenatbib]{mnras}

\usepackage{newtxtext,newtxmath} 

\usepackage[T1]{fontenc}
\usepackage{ae,aecompl}

\usepackage{graphicx}	
\usepackage{booktabs}
\usepackage{threeparttable}
\usepackage{subfig}
\usepackage{xcolor}
\usepackage{hyperref}
\hypersetup{colorlinks=true,linkcolor=blue,citecolor=blue,filecolor=blue,urlcolor=blue}

\graphicspath{{./}{image/}}

\newcommand{\phn}{\hphantom{0}}

\newcommand{\jgeod}{J.\ Geodesy} 
\newcommand{\jastata}{J.\ American Stat.\ Assoc.} 
\newcommand{\radsci}{Radio Sci.} 
\newcommand{\currsci}{Current Sci.} 

\title[Low-frequency observations of NGC~6251]{Low-frequency observations of the Giant Radio Galaxy NGC~6251}

\author[T. M. Cantwell et al.]{
T. M. Cantwell,$^{1}$ 
J. D. Bray,$^{1}$\thanks{E-mail: justin.bray@manchester.ac.uk}
J. H. Croston,$^{2}$
A.~M.~M.~Scaife,$^{1}$
D. D. Mulcahy,$^{1}$
\newauthor
P. N. Best,$^{3}$
M. Br\"uggen,$^{4}$
G. Brunetti,$^{5}$
J. R. Callingham,$^{6}$
A. O. Clarke,$^{1}$
\newauthor
M. J. Hardcastle,$^{7}$
J. J. Harwood,$^{7}$
G. Heald,$^{8}$
V. Heesen,$^{4,9}$
M. Iacobelli,$^{6}$
\newauthor
M. Jamrozy,$^{10}$
R. Morganti,$^{6,11}$
E. Orr\'u,$^{6}$
S. P. O'Sullivan,$^{4}$
C. J. Riseley,$^{8,5,12}$
\newauthor
H. J. A. R\"ottgering,$^{13}$
A. Shulevski,$^{14}$
S. S. Sridhar,$^{6}$
C. Tasse,$^{15,16}$
C. L. Van Eck$^{17}$
\\
$^{1}$ JBCA, Dept.\ of Physics \& Astronomy, University of Manchester, Manchester M13 9PL, UK\\
$^{2}$ School of Physical Sciences, The Open University, Walton Hall, Milton Keynes MK7 6AA, UK\\
$^{3}$ SUPA, Institute for Astronomy, Royal Observatory, Blackford Hill, Edinburgh, EH9 3HJ, UK\\
$^{4}$ Hamburger Sternwarte, University of Hamburg, Gojenbergsweg 112, 21029 Hamburg, Germany\\
$^{5}$ INAF --- Istituto di Radioastronomia, via P.\ Gobetti 101, 40129 Bologna, Italy\\
$^{6}$ ASTRON, the Netherlands Institute for Radio Astronomy, Postbus 2, 7990 AA, Dwingeloo, the Netherlands\\
$^{7}$ Centre for Astrophysics Research, School of Physics, Astronomy and Mathematics, University of Hertfordshire, Hertfordshire AL10 9AB, UK\\
$^{8}$ CSIRO Astronomy and Space Science, PO Box 1130, Bentley, WA 6102, Australia\\
$^{9}$ School of Physics and Astronomy, University of Southampton, Southampton SO17 1BJ, UK\\
$^{10}$ Astronomical Observatory, Jagiellonian University, ul.\ Orla 171, 30--244 Krakow, Poland\\
$^{11}$ Kapteyn Astronomical Institute, University of Groningen, P.O.~Box 800, 9700 AV Groningen, the Netherlands\\
$^{12}$ Dipartimento di Fisica e Astronomia, Universit\`a degli Studi di Bologna, via P.\ Gobetti 93/2, 40129 Bologna, Italy\\
$^{13}$ Leiden Observatory, Leiden University, PO Box 9513, NL-2300 RA Leiden, the Netherlands\\
$^{14}$ Anton Pannekoek Institute for Astronomy, University of Amsterdam, Postbus 94249, 1090 GE Amsterdam, the Netherlands\\
$^{15}$ GEPI, Observatoire de Paris, CNRS, Universite Paris Diderot, 5 place Jules Janssen, 92190 Meudon, France\\
$^{16}$ Dept.\ of Physics \& Electronics, Rhodes University, PO Box 94, Grahamstown, 6140, South Africa\\
$^{17}$ Dunlap Institute for Astronomy and Astrophysics, University of Toronto, 50 St.~George Street, Toronto, ON M5S 3H4, Canada
}

\date{Accepted XXX. Received YYY; in original form ZZZ}

\pubyear{2019}

\begin{document}
\label{firstpage}
\pagerange{\pageref{firstpage}--\pageref{lastpage}}
\maketitle

\begin{abstract}
We present LOFAR observations at 150\,MHz of the borderline FRI/FRII giant radio galaxy NGC~6251.  This paper presents the most sensitive and highest-resolution images of NGC~6251 at these frequencies to date, revealing for the first time a low-surface-brightness extension to the northern lobe, and a possible backflow associated with the southern lobe.  The integrated spectra of components of NGC~6251 are consistent with previous measurements at higher frequencies, similar to results from other LOFAR studies of nearby radio galaxies.  We find the outer structures of NGC~6251 to be either at equipartition or slightly electron dominated, similar to those of FRII sources rather than FRIs; but this conclusion remains tentative because of uncertainties associated with the geometry and the extrapolation of X-ray measurements to determine the external pressure distribution on the scale of the outer lobes.  We place lower limits on the ages of the extension of the northern lobe and the backflow of the southern lobe of $t \gtrsim 250$\,Myr and $t \gtrsim 210$\,Myr respectively.  We present the first detection of polarisation at 150\,MHz in NGC~6251.  Taking advantage of the high Faraday resolution of LOFAR, we place an upper limit on the magnetic field in the group of $B < 0.2 \, (\Lambda_B / 10\,{\rm kpc})^{-0.5}$\,$\mu$G for a coherence scale of $\Lambda_B < 60\,{\rm kpc}$ and $B < 13$\,$\mu$G for $\Lambda_B = 240$\,kpc.

\end{abstract}




\section{Introduction}

Giant radio galaxies (GRGs) are a population of radio galaxies with projected linear sizes greater than 1\,Mpc \mbox{\citep{willis1974}}.  These sources are typically found in galaxy groups, and in terms of their Fanaroff-Riley classification \citep[FR;][]{Fanaroff1974} are generally either FRII \citep[e.g.][]{shulevski2019} or borderline FRI/FRII \citep{Ishwara1999}, although examples of giants with FRI structure also exist \citep[e.g.][]{Heesen2018,dabhade2019}.  Due to their large physical extent, nearby GRGs allow detailed analysis of their jet and lobe structures \citep{Laing2006,Perley1984} as well as variations in the spectral index across the source \citep{Mack1997,Mack1998,Heesen2018}.

The origin of the Mpc sizes of GRGs has been investigated by many authors \citep{Komberg2009,Subrahmanyan2008,Machalski2004,Saripalli1997,Mack1998}.  GRGs are not thought to be intrinsically different from the more common smaller radio galaxies but rather a later stage in their evolution \citep{Machalski2006,Jamrozy2008,Komberg2009}.  \citeauthor{Machalski2006} argue that the correlation between the degree of depolarisation and the linear size suggests that the environments of GRGs also play a role in their formation.  X-ray observations of the intergalactic medium (IGM) of some GRGs combined with optical spectroscopic observations of the group galaxies show that the X-ray luminosity of the IGM is much lower, by as much as an order of magnitude, than would be expected from the correlation between X-ray luminosity and velocity dispersion \citep{Chen2011,Chen2012}, suggesting that the density of the environment is quite low.  However \citeauthor{Komberg2009} note that GRGs can be found in a range of environments ranging from very poor groups to clusters.

In many cases the lobes of GRGs appear to extend beyond their host environment into the large-scale structure (LSS) of the Universe.  Many GRGs exhibit asymmetries in their source structure, which may reflect asymmetries in their host environments \citep{Pirya2012,Schoenmakers2000,Lara2001}.  \citeauthor{Pirya2012} find that the shorter jet/lobe tends to be directed towards overdensities of galaxies.  The lobes of GRGs are potentially powerful indirect probes of the warm hot intergalactic medium (WHIM) that exists in large scale filaments.  The WHIM is a natural prediction of $\rm \Lambda$CDM cosmology and is thought to contain $\sim50\%$ of the baryonic matter in the Universe \citep{Dave2001,Nicastro2008,Smith2011}.  Recent Sunyaev-Zeldovich studies claim to have detected this low-density material for the first time \citep{degraaff2017,Tanimura2017}.  Indirect measurements of the WHIM using observations of GRGs provide an important complementary tool to trace this material, by assuming that the lobes of GRGs are relaxed and in equilibrium with the external WHIM pressure.  By calculating the internal pressure of the lobe we can therefore measure the pressure in the WHIM \citep{Subrahmanyan2008,Safouris2009}.  \citet{Malarecki2015} combine radio observations of GRGs with spectroscopic optical observations of nearby galaxies to demonstrate that it is possible to use GRGs to probe the denser regions of the WHIM. 

In order to calculate the internal pressure of the GRG~lobe it is necessary to make some assumptions about the particle energetics.  The simplest assumption one can make when calculating the internal pressure is that the relativistic electrons and magnetic field are in equipartition, with equal energy density \citep[e.g.][]{Hardcastle2002,Laing2002,Croston2004}.  This assumption can be tested for radio galaxies in those cases where X-ray observations are able to detect the intra-cluster medium or the inverse-Compton radiation of the lobes.  Such comparisons have been carried out for many sources.  In general it is found that FRII sources are close to equipartition, with high-energy electrons only slightly dominating over the energy of the magnetic field \citep[e.g][]{Brunetti1999,Hardcastle2000,Croston2005,Migliori2007,Isobe2015,Kawakatu2016,Ineson2017}.  In contrast, for FRI sources it is typically found that equipartition implies them to be significantly underpressured, with a significant violation of equipartition required for them to match the pressure of their surroundings \citep{Morganti1988,Worrall2000,Croston2008,Croston2014}.  The apparent difference in FRI and FRII particle content/energetics is discussed in detail by \citet{Croston2018}.

Past studies attempting to constrain the energetics in radio galaxies were limited by the lack of low-frequency observations.  The lobes of radio galaxies generally have steep spectra, and any variation from the assumed spectral behaviour at low frequencies could lead to large changes in the calculated energetics.  With the advent of new low-frequency instruments, such as the LOw Frequency ARray \citep[LOFAR;][]{vanHaarlem2013}, the recently-upgraded Giant Meterwave Radio Telescope \citep[GMRT/uGMRT;][]{swarup1991,gupta2017} and the Murchison Widefield Array \citep[MWA;][]{tingay2012}, we can now begin to constrain the behaviour of the low energy electron population.  Indeed recently \citet{Harwood2016} demonstrated that, in the case of FRII sources, the low-frequency spectra can be steeper than previously assumed, leading to an increase in the estimated total energy content of the lobes, as large as a factor of five in the case of 3C452.

In this paper we present total-intensity and polarised-intensity observations of the nearby GRG NGC~6251 at 150\,MHz with LOFAR high-band antennas (HBA).  NGC~6251 is a GRG with a projected linear size of $1.7$\,Mpc \citep{Perley1984} and a borderline FRI/FRII morphology.  The main jet and lobe are centre-brightened like an FRI; however, there is a hotspot or `warm spot' in the northern lobe suggestive of an FRII.  In contrast, the southern jet/lobe structure is edge-brightened, but possesses an inner hotspot somewhat reminiscent of wide-angle tail structures.  The radio power at 178\,MHz is $P_{\rm 178\,MHz}\approx1.4\times10^{25}$\,W\,Hz$^{-1}$ \citep{Waggett1977}, within an order of magnitude of the traditional \citet{Fanaroff1974} division between FRI and FRII sources ($\sim 10^{26}$\,W\,Hz$^{-1}$ in our assumed cosmology; see below) --- although note that this division is now known to be more blurred, and potentially strongly environmentally dependent \citep[e.g.][]{mingo2019}.  The large-scale morphology of NGC~6251 has some similarities with sources previously classed as ``hybrids'', but now thought to be strongly-projected sources with FRII-like jets \citep{harwood2019}; it is likely that projection as well as an intermediate jet power and environmental effects together explain the unusual structure.

There have been many radio observations of NGC~6251.  The first observations were carried out by \citet{Waggett1977} at 150\,MHz and 1.4\,GHz.  \citet{Perley1984} present detailed high-resolution Very Large Array (VLA) observations of the main jet in NGC~6251 at 1.4\,GHz.  \citet{Mack1997} and \citet{Mack1998} present observations of the large-scale structure of NGC~6251 from 325\,MHz to 10\,GHz.  Observations also show that NGC~6251 is highly linearly polarised, as much as 70\% in some regions, which is close to the theoretical maximum \citep{Willis1978,Stoffel1978,Saunders1981,Mack1997,Perley1984}.

X-ray observations have revealed an X-ray jet, as well as extended emission from the group-scale environment \citep{Mack1997c,Evans2005}.  \citeauthor{Evans2005} also used these observations to investigate the internal conditions in the lobes (but see the discussion in Section~\ref{sec:pressure}).  There have also been gamma-ray observations of NGC~6251: the \emph{Fermi} team reported detections of NGC~6251 as 1FGL J1635.4+8228 in the first-year \emph{Fermi} catalogue \citep{Abdo2010} and as 2FGL J1629.4+8236 in their second-year catalogue \citep{Nolan2012}.  The 95\% error on the position of 2FGL J1629.4+8236 includes both the jet and lobe of NGC~6251.  \citet{Takeuchi2012} observed NGC~6251 with Suzaku and detected diffuse X-ray emission in its northern lobe.  They argue that 2FGL J1629.4+8236 is consistent with non-thermal inverse-Compton emission from the lobes, based on detailed modelling of the spectral energy distribution (SED).

The aims of the work presented in this paper are two-fold.  Our first aim is to investigate the low-frequency radio-continuum spectral behaviour of NGC~6251 and re-examine the pressure balance in its lobes, taking into account the new LOFAR data.  Our second aim is to probe the environment and source structure using the high-resolution Faraday spectra obtained using the LOFAR HBA data.  The material in this paper is split between Section~\ref{sec:obs_data_Red}, in which we describe the observational data and its basic processing, Section~\ref{sec:results}, in which we derive results regarding spectra and polarisation, Section~\ref{sec:discussion}, in which we discuss these results with reference to the above aims, and Section~\ref{sec:conclusion}, in which we summarise our conclusions.

A preliminary report of this work has been previously published \citep{Therese_thesis}, and contains additional details of some intermediate results that are omitted here for clarity.  This paper, with the benefit of peer review, confirms the main conclusions of the preliminary report, but more rigorously defines the conditions under which they are valid, and extends them: in particular, it extends the limit on the magnetic field in the group environment out to larger coherence scales (see Section~\ref{sec:Blim}).

In this work, a $\Lambda$CDM cosmology is assumed with $H_{0}=70\rm \, km\, s^{-1}\,Mpc^{-1}$, $\Omega_{m}=0.3$, $\Omega_{\Lambda}=0.7$.  Using these parameters, at a redshift of 0.02471, 1\,arcsec corresponds to a physical scale of 0.498\,kpc \citep{Wright2006}.  Spectral indices $\alpha$ are defined in the sense $S_{\nu}\propto\nu^{\alpha}$.

\section{Observations and imaging}
\label{sec:obs_data_Red}

NGC~6251 was observed with LOFAR HBA on 23~August 2013 during LOFAR's cycle~0.  A summary of the observations is provided in Table~\ref{Table:Obs_dets}.  Data were taken in interleaved mode, with scans alternating between the target and the flux calibrator.  This mode was used in early LOFAR cycles
to compensate for gain instability, before other approaches were developed.  The calibrator 3C295 was observed for 2~minutes per scan and the target scans were 10~minutes long.

\begin{table}
 \centering
  \caption{Observation details for NGC~6251.}
  \label{Table:Obs_dets}
  \begin{tabular}{@{}ll@{}}
  \toprule
Project code & $\rm LC0\_012$\\
Date & 2013-Aug-23\\
Central frequency (MHz) & 150\\
Time on target (h) & 6.5\\
Bandwidth (MHz) & 80\\
Usable bandwidth (MHz) & 63\\
Channels/sub-band & 64\\
Averaged channels/sub-band & 4\\
\% flagged & 38\%\\
Sensitivity & 2\,mJy\,beam$^{-1}$\\
Angular resolution & 40\,arcsec\\
FOV & $\sim 6\times6$\,deg$^2$\\
\bottomrule
\end{tabular}
\end{table}

\subsection{LOFAR}
\subsubsection{Calibration and imaging}
\label{sec:HBA-cal-imaging}

An initial flagging step was performed using \textsc{AOFlagger} \citep{offringa2012}\footnote{ \url{https://sourceforge.net/projects/aoflagger}}.  3C295 was then calibrated using BlackBoard SelfCal (\textsc{bbs}) and a simple two-component model.  The flux scale was set using \citet{Scaife2012}.  These solutions were transferred to the target and then a phase-only self-calibration was performed on each sub-band using the LOFAR global skymodel (gsm) \citep{Smirnov2004}.  The data were imaged with \textsc{AWimager} \citep{Tasse2013} to generate a new sky model, which was then used to perform a single round of phase-only self-calibration.  The data were combined into 18~bands of 3.515\,MHz each with a channel width of 48\,kHz.  We did not carry out direction-dependent calibration \citep[e.g.][]{vanWeeren2016}, as our primary target was the bright central source, and the image quality achievable with a direction-independent calibration was sufficient for our science aims.

\begin{figure*}
  \centering
     \includegraphics[width=1\textwidth]{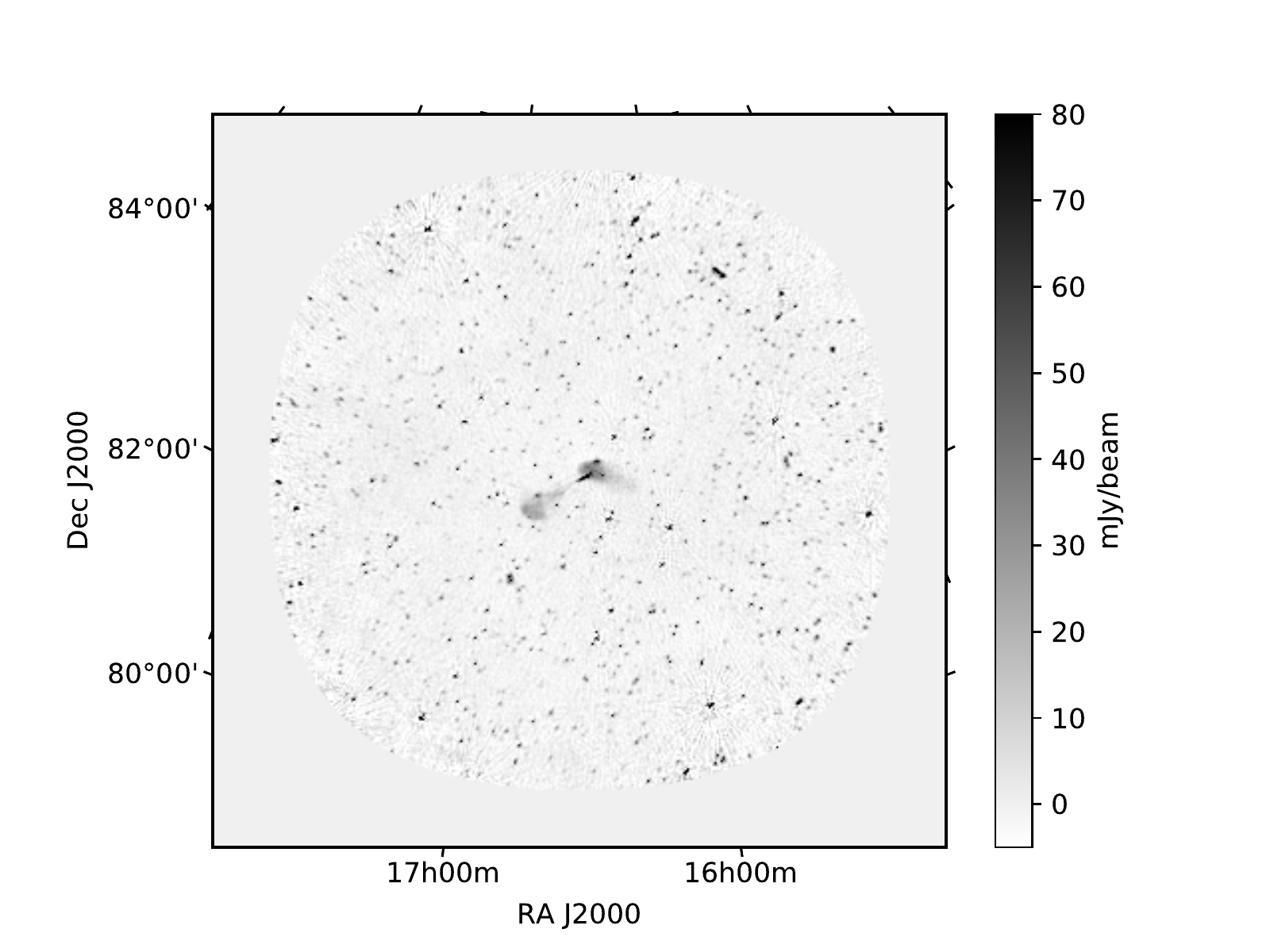}
     \caption{Greyscale image showing the full primary-beam-corrected LOFAR HBA field of view total-intensity map.}
     \label{fig:HBA_FOV}
\end{figure*}

\begin{figure*}
  \centering
     \includegraphics[width=1.2\textwidth]{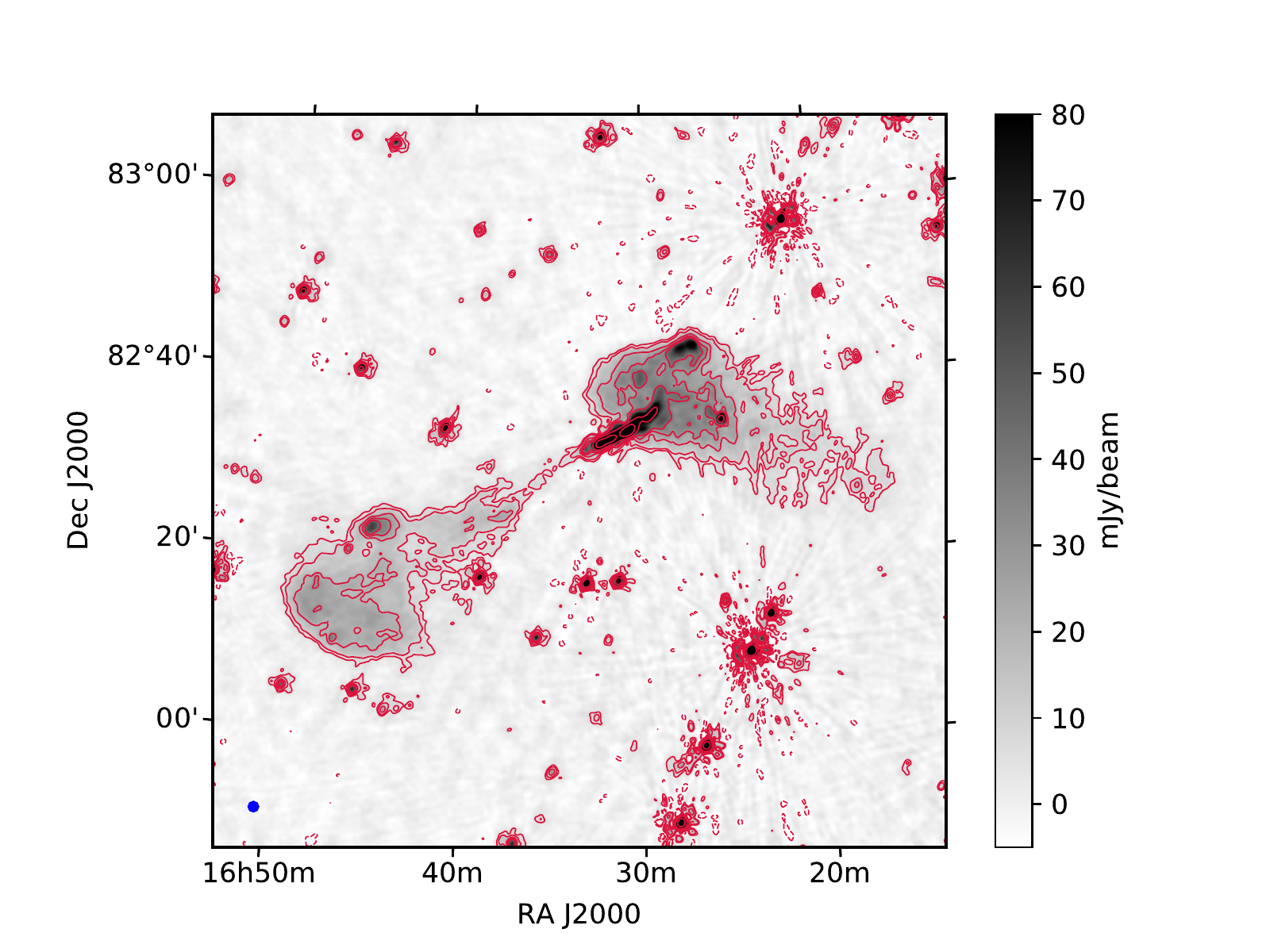}
     \caption{Contours and greyscale image showing the LOFAR 150\,MHz HBA total-intensity map of NGC~6251.  Contours are shown at -3, 3, 5, 10, 15, 20, 60, 150, 200, 400~$\times$~$\sigma_{\rm rms}$ where $\sigma_{\rm rms}=2.0$\,mJy\,beam$^{-1}$.  The blue circle in the bottom left-hand corner shows the beam resolution.  Radial artefacts are visible around background sources to the northwest and southwest, but fall below the first contour threshold well before they reach the northern lobe of NGC~6251, so we do not expect them to appreciably affect our results.}
     \label{fig:HBA_total_Intensity}
\end{figure*}

Final imaging was carried out using \textsc{AWImager} using Briggs weighting with a robustness parameter of~0.  Due to issues with radio-frequency interference (RFI), only 63\,MHz of the 80\,MHz total bandwidth was used.  In order to investigate the diffuse emission, each band was imaged separately, with an outer \textit{uv}~limit of 3k$\lambda$ in units of wavelength~$\lambda$, achieving a resolution of 40\,arcsec.  At this point the flux scale was corrected as described in Section~\ref{sec:hba_fluxscale}.  The flux-corrected images were combined to produce a weighted average
 $ \bar{X} = \frac{ \sum_i{ \sigma_{{\rm rms,}i}^{-2} \, X_i } }{ \sum_i \sigma_{{\rm rms},i}^{-2} } $
where $X_i$ is the image in band $i$ and $\sigma_{{\rm rms,} i}$ is its root-mean-square (rms) noise.  The effective frequency for this weighted average is 140\,MHz, though throughout the rest of this document we label these data with the nominal 150\,MHz midpoint frequency for this LOFAR band.  Fig.~\ref{fig:HBA_FOV} shows the full field of view, and Fig.~\ref{fig:HBA_total_Intensity} shows a zoomed-in image of NGC~6251.  The expected thermal noise with these parameters is approximately $0.2$\,mJy\,beam$^{-1}$, while the measured noise in our images is 2\,mJy\,beam$^{-1}$.  This increase in noise is typical for data which, like ours, have not undergone direction-dependent calibration \citep{vanWeeren2016}. 

\subsubsection{Flux-density scale}
\label{sec:hba_fluxscale}

There are known problems with the LOFAR HBA flux-density scale \citep{Heald2015,Hardcastle2016}.  As with any aperture array, the LOFAR primary beam is elevation--dependent, leading to different primary beams when observing the calibrator source and the target source.  This difference should be accounted for when transferring the amplitude gains from the calibrator source to the target field.  However, as the overall normalisation of the LOFAR HBA beam is poorly constrained, it is not currently possible to include this effect directly during calibration.  This leads to a  frequency-dependent effect on the LOFAR HBA fluxes.  In order to correct for this effect, we follow the flux boot-strapping procedure outlined by \citet{Hardcastle2016}\footnote{\url{https://github.com/mhardcastle/lofar-bootstrap}}.

First a catalogue of sources was generated for our LOFAR field using \textsc{pybdsf}\footnote{\textsc{pybdsf} documentation: \url{http://www.astron.nl/citt/pybdsm/}}.  From this catalogue, bright sources with fluxes $>0.1$Jy were cross-matched with the VLA Low-Frequency Sky Survey \citep[VLSSr;][]{Lane2012} and the NRAO VLA Sky Survey \citep[NVSS;][]{Condon1998}.  The final catalogue of sources contained only those with sources having both a VLSSr counterpart and an NVSS counterpart.  A flux correction factor was then found for each band and applied to the LOFAR field, and a new source catalogue for the field was generated.

To test the reliability of this flux correction, we found the spectral index for every source within the half-power distance of the LOFAR primary beam with an NVSS counterpoint.  The spectral indices of these sources have a mean of $-0.8$ and standard deviation of $0.3$.  This is consistent with our expectation, suggesting that our corrected flux scale is reliable.


\subsubsection{Polarisation imaging}
\label{sec:polimg}

The commissioning of a pipeline to process LOFAR polarisation data is not yet complete, and so it is not currently possible to calibrate the instrumental polarisation and thus to determine the absolute polarisation angle.  However, it is still possible to detect polarised emission with LOFAR \citep{Mulcahy2014,osullivan2019}, and science commissioning has shown that linearly-polarised intensity and Faraday-depth values can be reliably recovered for known polarised sources.  The Faraday depth is defined as
\begin{equation}
\phi = 0.81\int{n_{\rm e}B\cdot dl} \textrm{\,rad\,m$^{-2}$},
\label{eqn:Faraday}
\end{equation}
where $n_{\rm e}$ is the electron density in cm$^{-3}$, $B$ the magnetic field in $\mu$G, and $dl$ the path length in parsec.

\begin{figure}
 \centering
 \includegraphics[width=0.5\textwidth]{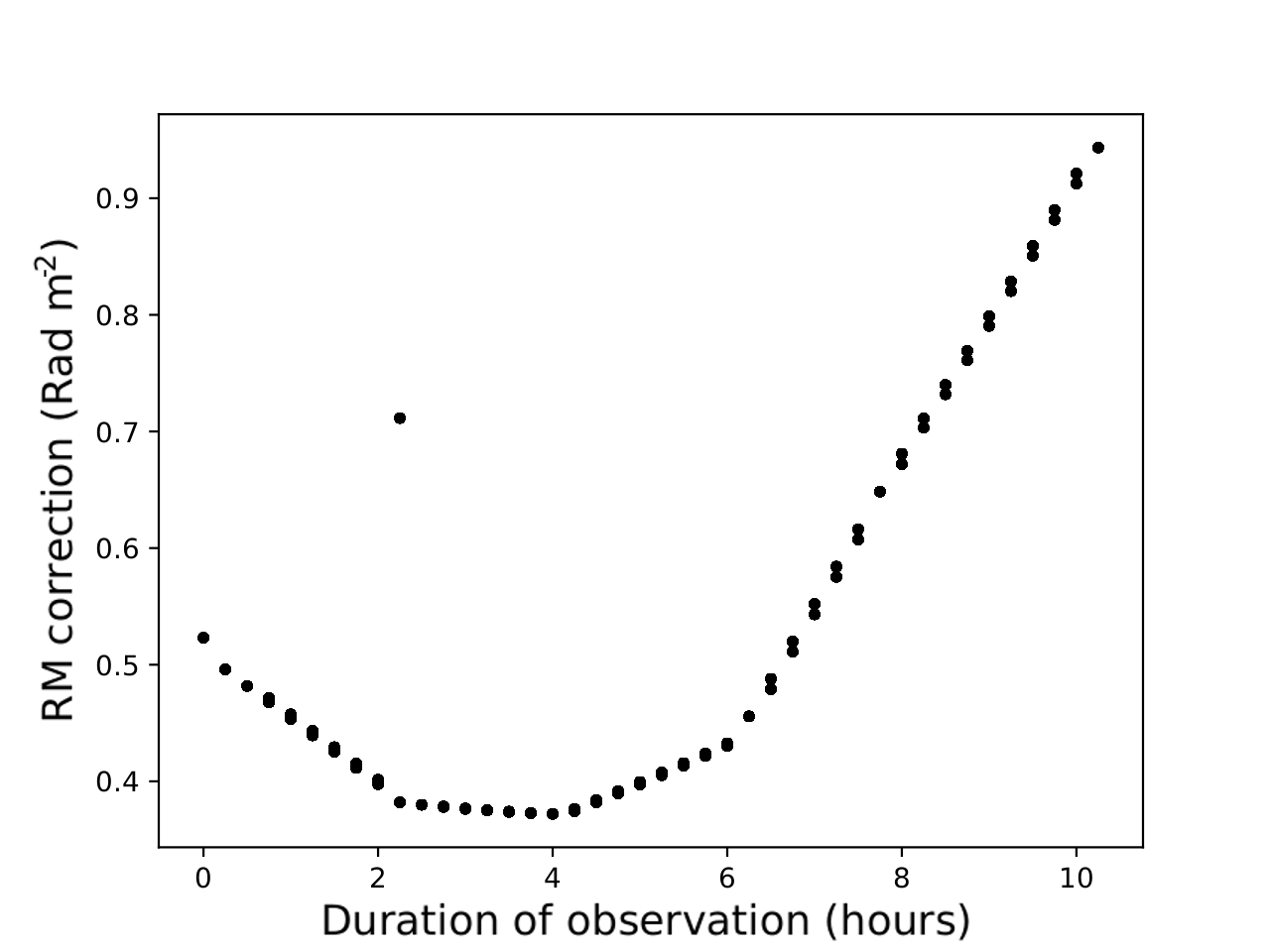}
  \caption{Faraday rotation due to the ionosphere.  There is a single outlier visible in the plot, which was excluded from our analysis.}
  \label{fig:rm_correction}
\end{figure}

To image polarised emission from NGC~6251 we must account for Faraday rotation due to the ionosphere, which results, per equation~(\ref{eqn:Faraday}), from ionospheric free electrons and the geomagnetic field.  Variations in the ionospheric electron content and the projection angle of the geomagnetic field during the observations will lead to different degrees of Faraday rotation throughout the data, causing a smearing of any signal in Faraday space \citep{sotomayor2013}.  We corrected for ionospheric Faraday rotation with \textsc{RMExtract}\footnote{\url{https://github.com/maaijke/RMextract}} \citep{Maaijke2018}, which calculates the expected Faraday rotation over the LOFAR stations from a model of the geomagnetic field and maps of the ionospheric total electron content (TEC).  The geomagnetic field is taken from the International Geomagnetic Reference Field (IGRF), and the TEC maps may be obtained from either the Centre for Orbital Determination in Europe (CODE)\footnote{\url{http://aiuws.unibe.ch/ionosphere/}} or the Royal Observatory of Belgium (ROB)\footnote{\url{http://gnss.be/Atmospheric_Maps/ionospheric_maps.php}}.  Tests during commissioning investigating pulsars of known properties suggest that using CODE ionospheric maps recovers more of the true polarised flux, and so we used these as the input for the ionospheric correction \citep{vanEck}.  CODE calculates the TEC using data from $\sim$200~GPS and Global Navigation Satellite System (GLONASS) sites of the International GPS Service (IGS) and other institutions, with a time resolution of about an hour, and a spatial resolution of $2.5^{\circ}\times5.0^{\circ}$ \citep{Dow2009}.  Fig.~\ref{fig:rm_correction} shows a plot of the ionospheric rotation-measure (RM) correction produced by RMExtract.

Once the RM~correction had been applied to every sub-band and time step, individual channels of 48\,kHz were split from sub-bands and imaged in Stokes~Q and~U using \textsc{AWImager}.  An inner \textit{uv}~limit of $200\lambda$, corresponding to an angular scale of $\sim 20$\,arcmin, was used in order to avoid imaging Galactic foreground emission.

\subsection{Archival data}

We have used a number of archival datasets in our analysis of NGC~6251.  We have used the Westerbork Synthesis Radio Telescope (WSRT) 325\,MHz and 610\,MHz images as well as the Effelsberg 10\,GHz images from \citet{Mack1997}, which are discussed in detail in that work and by \citet{Mack1998}.  A number of VLA datasets from the archive were also used; the details of these are summarised in Table~\ref{Table:archival_vla_data}.  The VLA datasets used were chosen to best match the resolution of the LOFAR observations.  Observations at 8\,GHz in D~configuration as well as 1.4\,GHz and 325\,MHz in B~configuration were used to image the core of NGC~6251.  Observations at 1.4\,GHz in D~configuration were used to analyse the large-scale structure of NGC~6251. 

\begin{table*}
 \centering
  \caption{Details of archival VLA data used in this work, including the configuration of the telescope at the time of the observations, the frequency used, and the reference for the image based on the data.}
  \label{Table:archival_vla_data}
  \begin{tabular}{@{}lllll@{}}
  \toprule
Proposal ID & Date & Configuration & Frequency & Reference \\
\midrule
AK461     & 5-Oct-1998  & B & 325\,MHz & This work\\
VJ49,VJ38 & 20-Nov-1988 & A,B & 1.4\,GHz & \citet{Evans2005}\\
Test      & 5-Dec-1985  & D & 1.4\,GHz & This work\\
AB3346    & 1-Dec-1985  & D & 1.4\,GHz & This work\\
AM0322    & 9-May-1991  & D & 8\,GHz & \citet{Evans2005}\\
\bottomrule
\end{tabular}
\end{table*}

\begin{figure}
  \centering
     \includegraphics[width=0.5\textwidth]{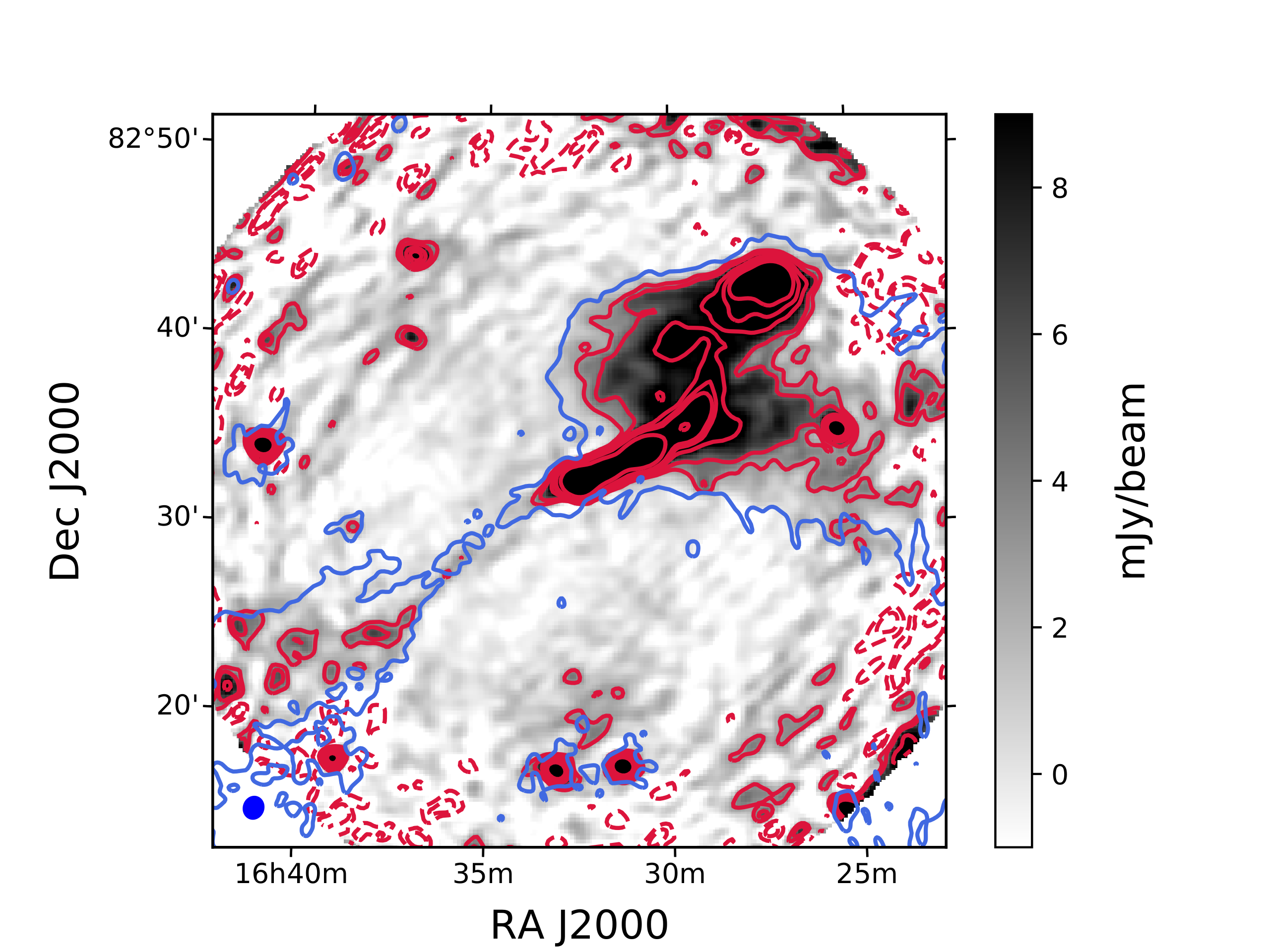}
     \caption{Northern lobe of NGC~6251.  Greyscale shows the VLA 1.4\,GHz D-configuration image, with contours in red at -5, -3, 5, 10, 20, 30, 40, 100~$\times$~$\sigma_{\rm rms}$ where $\sigma_{\rm rms}=1.0$\,mJy\,beam$^{-1}$.  LOFAR 150\,MHz HBA contours are shown in blue at $3\sigma_{\rm rms}$ where $\sigma_{\rm rms}=2.$0\,mJy\,beam$^{-1}$.  The blue circle in the bottom left-hand corner shows the VLA beam resolution.}
     \label{fig:vla_north_images}
\end{figure}

\begin{figure}
  \centering
     \includegraphics[width=0.5\textwidth]{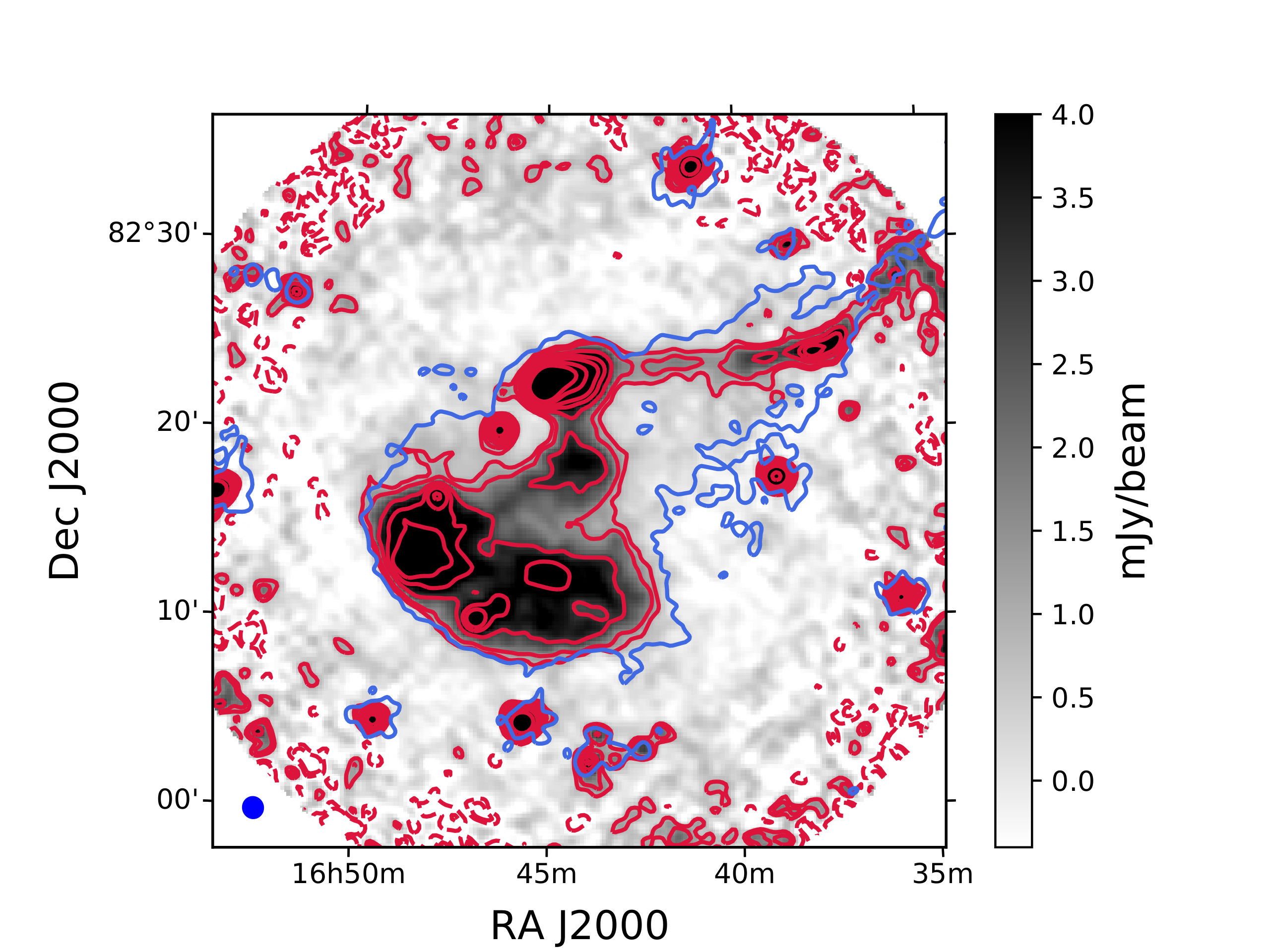}
     \caption{Southern lobe of NGC~6251.  VLA 1.4\,GHz D-configuration and LOFAR 150\,MHz HBA data are shown as in Fig.~\ref{fig:vla_north_images}.} \label{fig:vla_south_images}
\end{figure}

The B-configuration 325\,MHz data and D-configuration 1.4\,GHz data were imaged and reduced in \textsc{casa\,4.7} \citep{McMullin2007}.  A simple calibration strategy was adopted for the D-configuration 1.4\,GHz data.  The flux scale was that of \citet{Perley2013}.  An initial phase calibration was performed using the flux calibrator followed by the bandpass calibration and a final amplitude and phase calibration.  NGC~6251 was observed as two pointings, one centred on the core and the other on the southern lobe.  Both pointings were imaged in two steps.  In the first round of imaging we applied a mask that excluded large extended regions, and did not carry out multi-scale cleaning.  Once all compact emission or narrow emission, such as the jet, was included in the model, a second round of imaging was carried out using multi-scale \textsc{clean}, in order to properly image the diffuse emission.  The data were imaged with a \textit{uv}~range of 140--4400$\lambda$ and natural weighting.  The final images of the northern and southern lobes are shown in Figs~\ref{fig:vla_north_images} and~\ref{fig:vla_south_images} respectively.

\begin{figure}
  \centering
     \includegraphics[width=0.5\textwidth]{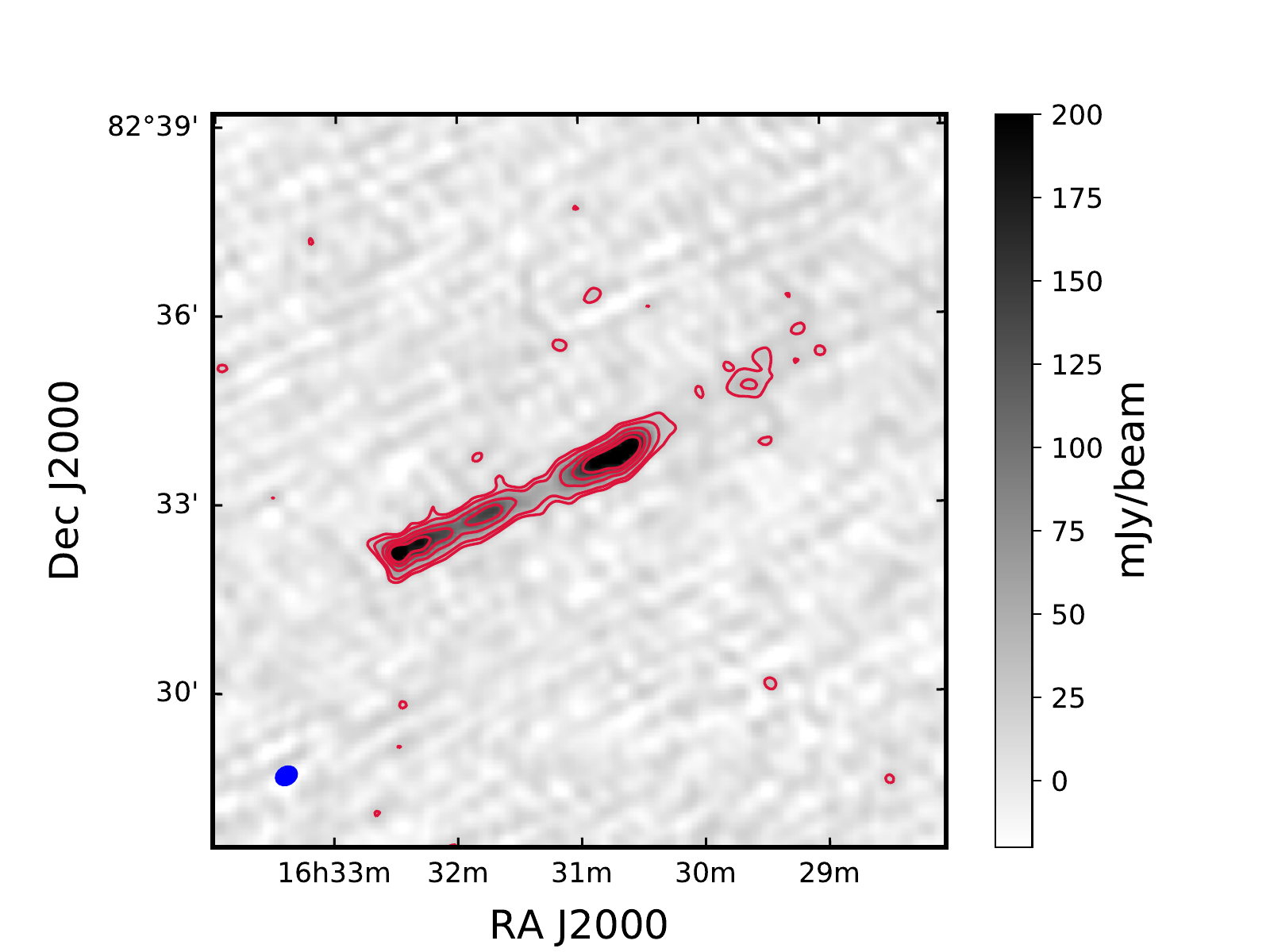}
     \caption{Contours and greyscale image showing the VLA 325\,MHz B-configuration map of NGC~6251.  Due to the high resolution, extended emission is resolved out; this figure shows only the inner region of the image where compact structure is visible.  Contours are shown at -3, 3, 5, 10, 15, 20~$\times$~$\sigma_{\rm rms}$ where $\sigma_{\rm rms}=7.0$\,mJy\,beam$^{-1}$.  The blue circle in the bottom left-hand corner shows the 
     beam resolution.}
     \label{fig:vla_P-band}
\end{figure}

The VLA 325\,MHz data were calibrated similarly, with one additional step at the start of the procedure.  Data from the Global Positioning System (GPS) were used to generate a map of the ionospheric electron content, and a phase correction based on this map was applied to the data using the \textsc{casa} task {\it gencal}.
The resulting image is shown in Fig.~\ref{fig:vla_P-band}.  The resolution and rms noise of this and the other images used in this paper are summarised in Table~\ref{Table:image_sum}.

\begin{table*}
 \centering
  \caption{Summary of NGC~6251 images used in this work.}
  \label{Table:image_sum}
  \begin{tabular}{@{}lllll@{}}
  \toprule
Array & Frequency & Resolution (arcsec) & $\sigma_{\rm rms}$ (mJy\,beam$^{-1}$) & Reference\\
\midrule
LOFAR HBA          & 150\,MHz & 40 & 2.0 & This paper\\
WSRT               & 325\,MHz & 55 & 2.0 & \citet{Mack1997} \\
VLA B~config.      & 325\,MHz & 20 & 7.0 & This paper\\
WSRT               & 610\,MHz & 28 & 0.4 & \citet{Mack1997} \\
VLA D~config.\ (North Lobe) & 1.4\,GHz & 58 & 1.0 & This paper\\
VLA D~config.\ (South Lobe) & 1.4\,GHz & 55 & 0.3 & This paper\\
Effelsberg         & 10\,GHz  & 69 & 1.0 & \citet{Mack1997}\\
\bottomrule
\end{tabular}
\end{table*}

\section{Results}
\label{sec:results}

The full-bandwidth LOFAR image of NGC~6251 is shown in Fig.~\ref{fig:HBA_total_Intensity}.  The main jet extends north-west from the core, with a bright knot at a distance of 200\,arcsec (or 99.6\,kpc in projection).  The jet then bends north, and terminates at a hotspot.  The northern lobe overlaps with the hotspot and jet down as far as the knot, but a diffuse, low-surface-brightness component extends west from the lobe.  This extension was detected in the 325\,MHz map of \citet{Mack1997}; however, the 150\,MHz LOFAR data presented here show that the region extends at least a further 14.4\,arcmin (or 430\,kpc), so that the total length of the faint extension is 19\,arcmin (or 581\,kpc).

The counterjet is detected at a 3$\sigma$ level in the LOFAR image shown in Fig.~\ref{fig:HBA_total_Intensity}, which is the clearest detection of the counterjet at these frequencies to date.  The counterjet extends to the south-east.  At 700\,arcsec (or 349\,kpc in projection) from the core, the jet bends to the east.  The bend is bright and detected at 325\,MHz, 610\,MHz and 1.4\,GHz.  The VLA 1.4\,GHz image in Fig.~\ref{fig:vla_south_images} shows that the brightened jet continues eastward in a linear fashion until it reaches a bright, compact hotspot.  The jet is again deflected at the hotspot and continues to the south-east before terminating in a well-defined southern lobe. 

A region of diffuse low-surface-brightness emission can be seen coincident with the southern jet.  This emission was previously only seen in the 150\,MHz map of \citet{Waggett1977}.  As such this appears to be very steep-spectrum emission, and may originate in lobe material that has been deflected back towards the core.  We henceforth refer to this region as the `southern backflow'; for further discussion, see Section~\ref{sec:specage}.

\begin{figure*}
  \centering
     \includegraphics[width=\textwidth]{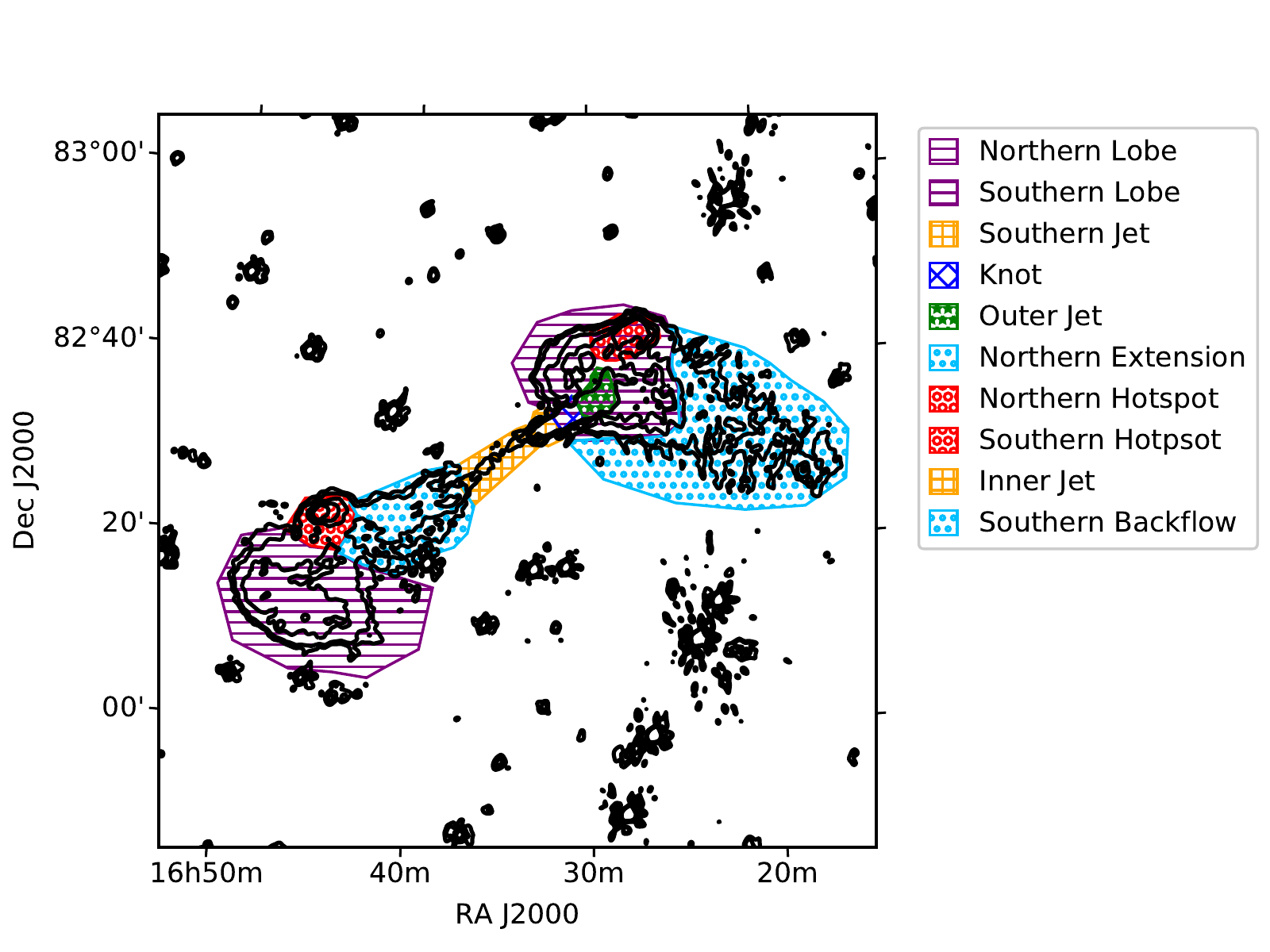}
     \caption{Illustration of the regions we define for each component of NGC~6251, used to integrate intensities to obtain flux densities for those components.  Contours shown are a subset of those in the LOFAR 150\,MHz image in Fig.~\ref{fig:HBA_total_Intensity}.}
     \label{fig:NGC6251_regions}
\end{figure*}

Table~\ref{tab:flux} shows the flux densities measured within the 3$\sigma$~contour line for individual components of NGC~6251, for a range of frequencies between 150\,MHz and 10\,GHz.  Fig.~\ref{fig:NGC6251_regions} shows a map of the regions used to define the components.  Point sources embedded in the lobe emission were replaced with blanked pixels.  Errors in the flux measurements were calculated using the equation
 \begin{equation} \label{eq:fluxerror}
  \sigma_{ S_{\rm \nu}}=\sqrt{\left(\sigma_{\rm cal}S_{\nu}\right)^{2}+\left(\sigma_{\rm rms}\sqrt{N_{\rm beam}}\right)^2}
\end{equation}
where $N_{\rm beam}$ is the number of independent beams in the region and $\sigma_{\rm cal}$ is the fractional uncertainty in the calibration of the flux-scale, which we take to be 10\%.  For the southern jet, we placed $3\sigma$~upper limits on the flux density at 325\,MHz, 610\,MHz and 1.4\,GHz assuming the surface brightness at these frequencies to be uniform over the region in which the jet was detected in the 150\,MHz image.  We placed similar limits at 325\,MHz and 610\,MHz for the northern extension, but not at 1.4\,GHz, as the VLA 1.4\,GHz data did not cover this area.

\begin{table*}
  \centering
  \begin{threeparttable}[b]
  \caption{Measured flux densities and limits for individual components of NGC~6251 at various frequencies from 150\,MHz to 10\,GHz, and fitted spectral indices $\alpha$ (see Section~\ref{sec:alpha}).  Flux-density data are from the images listed in Table~\ref{Table:image_sum}, except for the 8\,GHz value, which is from \citet{Evans2005}.}
  \label{tab:flux}
  \begin{tabular}{@{}llllllll@{}}
    \toprule
    Component & \multicolumn{6}{c}{Flux density (Jy)} & \multicolumn{1}{c}{$\alpha$} \\
    \cmidrule{2-7}
     & $S_{\rm 150\,MHz}$ & $S_{\rm 325\,MHz}$ & $S_{\rm 610\,MHz}$  & $S_{\rm 1.4\,GHz}$ & $S_{\rm 8\,GHz}$ & $S_{\rm 10\,GHz}$ & \\
    \midrule
Core Region             & $0.2\pm0.1$\tnote{$\star$} & $0.27\pm0.04$\tnote{$\dagger$} & $0.32\pm0.07$\tnote{$\star$} & $0.4\pm0.1$ & $0.72\pm0.03$ & $0.8\pm0.3$\tnote{$\star$} & $0.3\pm0.1$ \\
Inner Jet (core incl.)        & $2.2\pm0.2$   & $1.4\pm0.1$     & $1.2\pm0.1$     & $0.9\pm0.1$ & --- & $0.81\pm0.08$ & --- \\
Inner Jet (core excl.)        & $1.9\pm0.2$   & $1.2\pm0.1$     & $0.9\pm0.1$     & $0.5\pm0.1$ & --- & $0.1\pm0.3$ & $-0.6\pm0.2$ \\
Knot               				  & $2.8\pm0.3$   & $1.7\pm0.2$     & $1.2\pm0.1$     & $0.75\pm0.08$ & --- & $0.22\pm0.02$ & $-0.60\pm0.07$ \\
Outer Jet                         & $1.8\pm0.2$   & $0.90\pm0.09$   & $0.56\pm0.06$   & $0.34\pm0.04$ & --- & $0.072\pm0.008$ & $-0.75\pm0.08$ \\
Northern Lobe                     & $6\pm1$       & $2.0\pm0.5$     & $1.0\pm0.3$     & $0.5\pm0.1$   & --- & --- & $-1.1\pm0.3$ \\
Northern Extension                & $2.6\pm0.3$   & $<0.32$         & $<0.28$         & ---             & --- & --- & $<-2.7$\tnote{$\ddagger$} \\
Northern Hotspot                  & $2.0\pm0.2$   & $0.93\pm0.09$   & $0.59\pm0.06$   & $0.35\pm0.04$ & --- & $0.083\pm0.009$ & $-0.73\pm0.08$ \\
Southern Jet 					  & $0.26\pm0.03$ & $<0.09$         & $<0.08$         & $<0.03$      & --- & --- & $<-1.4$\tnote{$\ddagger$} \\ 
Southern Backflow                 & $1.8\pm0.2$    & $<0.169$        & $<0.1$          & $0.05\pm0.01$ & --- & --- & $-1.6\pm0.2$ \\
Southern Lobe                     & $5.8\pm0.6$   & $1.5\pm0.2$     & $0.51\pm0.05$   & $0.44\pm0.06$ & --- & --- & $-1.3\pm0.2$ \\
Southern Hotspot                  & $1.0\pm0.1$   & $0.30\pm0.03$   & $0.20\pm0.02$   & $0.15\pm0.02$ & --- & $0.022\pm0.003$ & $-0.85\pm0.09$ \\
   \bottomrule
 \end{tabular}
  \begin{tablenotes}
 \item [$\star$] Predicted using core spectral index calculated from VLA 325\,MHz, 1.4\,GHz and 8\,GHz data.
 \item [$\dagger$] Measured from VLA 325\,MHz data rather than the WSRT image at the same frequency.
 \item [$\ddagger$] Limit calculated from LOFAR 150\,MHz flux density and WSRT upper limits at 325\,MHz and 610\,MHz.
\end{tablenotes}
\end{threeparttable}
 \end{table*}

\subsection{Spectral index}
\label{sec:alpha}

\begin{figure*}
  \centering
  \begin{tabular}{cc}
     \subfloat[][]{\label{fig:northern_regions}}{\includegraphics[width=0.5\textwidth]{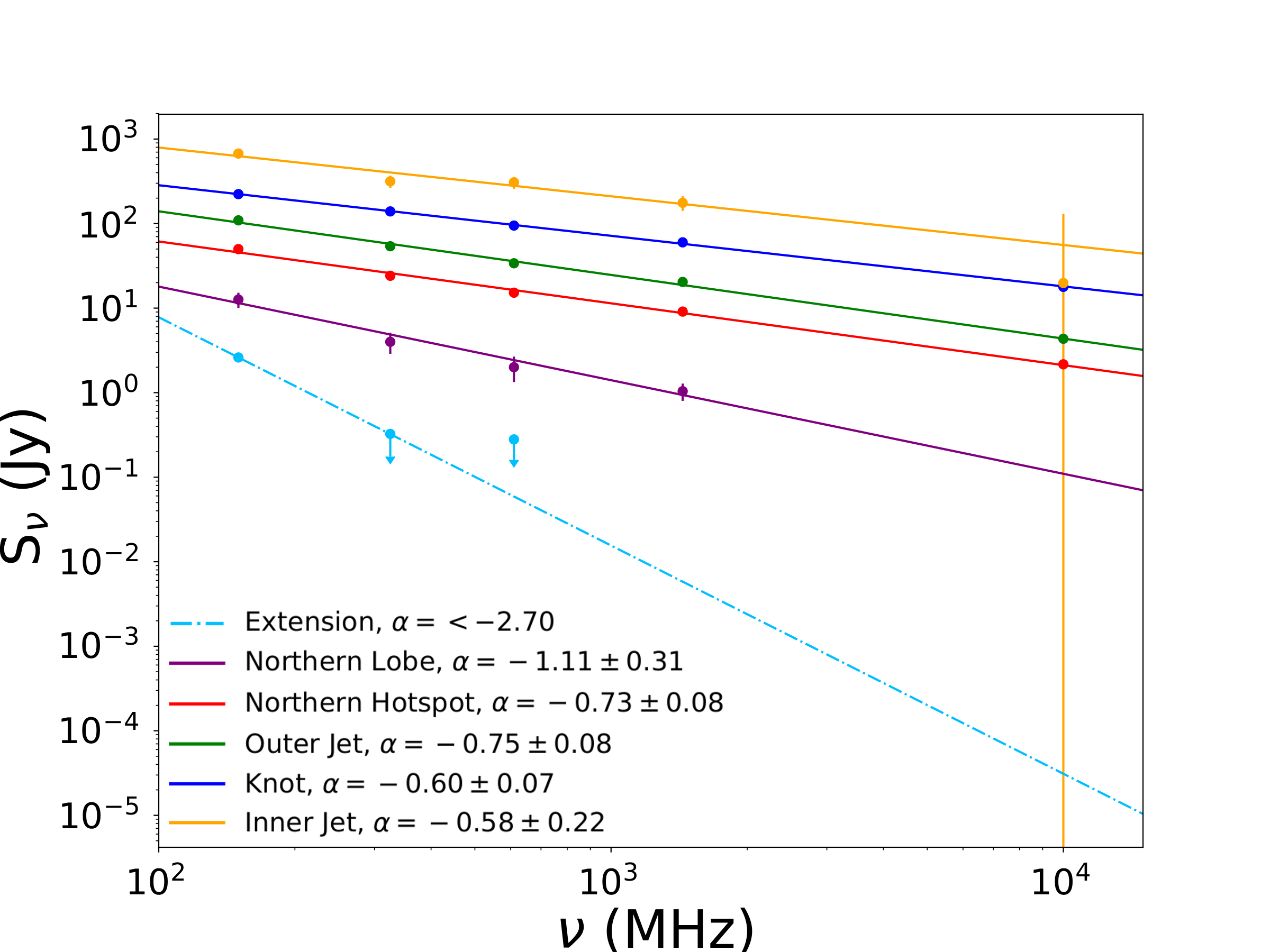}} & \subfloat[][]{\label{fig:knot}}{\includegraphics[width=0.5\textwidth]{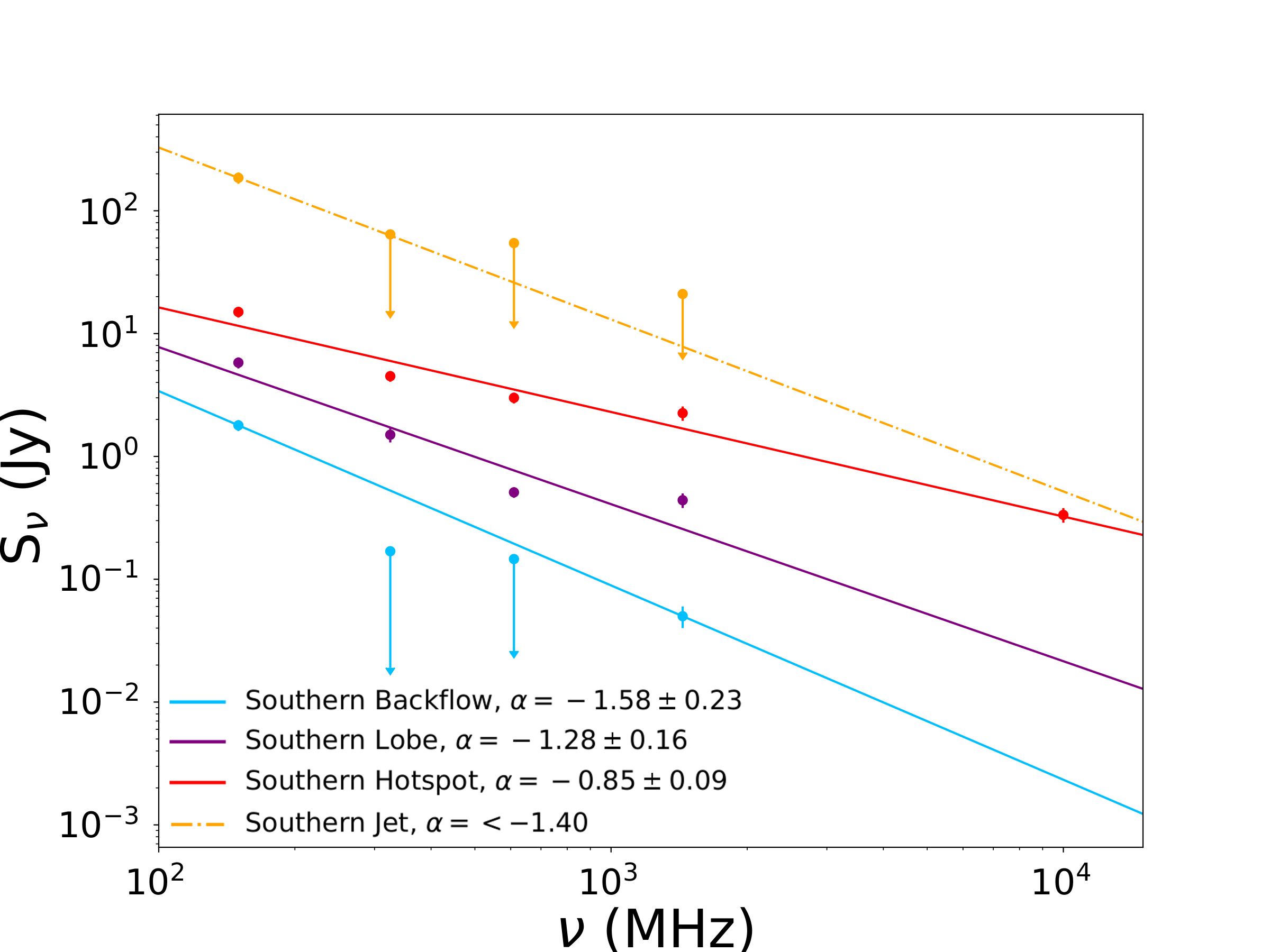}} \\
 \end{tabular}
     \caption{Integrated spectra and power-law fits for each component of NGC~6251, as defined in Fig.~\ref{fig:NGC6251_regions}, except the core region.  Fluxes are arbitrarily scaled to fit on the plot.  Note that the emission from the southern backflow is likely to come from multiple components, e.g.\ jet emission contributes at 1.4\,GHz.  Similarly, several points at 1.4\,GHz may include confusion noise from the bright core, which would explain the unusual convex spectra.  Dashed lines represent upper limits on the spectrum calculated from the LOFAR data and WSRT 325\,MHz 3$\sigma$ upper limits.}
     \label{fig:spectral_index_plots}
\end{figure*}

The integrated spectral index was calculated for each component of NGC~6251 using the fluxes shown in Table~\ref{tab:flux}.  Fig.~\ref{fig:spectral_index_plots} shows the best-fitting power law for the spectrum of each component and Table~\ref{tab:flux} lists the fitted spectral indices.  It should be noted that the \textit{uv}~range of the interferometric maps used to measure the flux densities are not matched as we did not have access to the \textit{uv}~data for all the images.  This could lead to an artificial steepening of the measured spectral index in regions of diffuse extended emission such as the lobes.

The core of NGC~6251 can not be separated from the inner jet region in the LOFAR, WSRT or low-resolution 1.4\,GHz images.  The flux from the core contributes to the inner jet region.  In order to subtract the core contribution from the inner jet region, archival 325\,MHz, 1.4\,GHz and 8\,GHz VLA data were used.  Table~\ref{tab:flux} shows the core fluxes measured from each of these datasets.  The spectral index of the core as measured from these data is inverted and has a value of $\alpha=+0.3\pm0.1$.  Using this spectral index, the core flux was predicted for each of our datasets and subtracted from the integrated flux of the inner jet region.

\begin{figure*}
  \centering
  \begin{tabular}{cc}
     \subfloat[][]{\label{fig:VLA_north_spix}}{\includegraphics[width=0.5\textwidth]{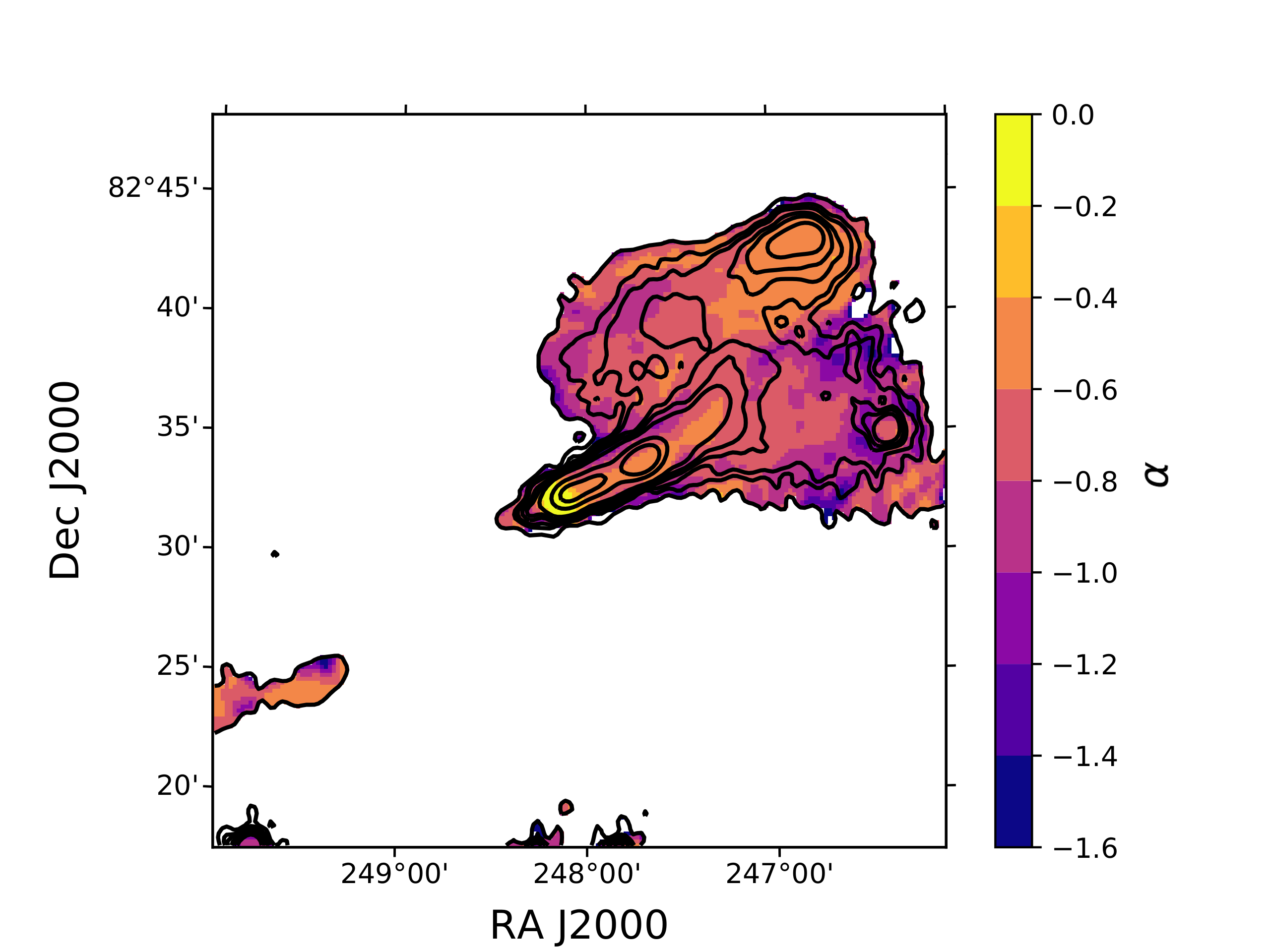}} & \subfloat[][]{\label{fig:VLA_north_spix_error}}{\includegraphics[width=0.5\textwidth]{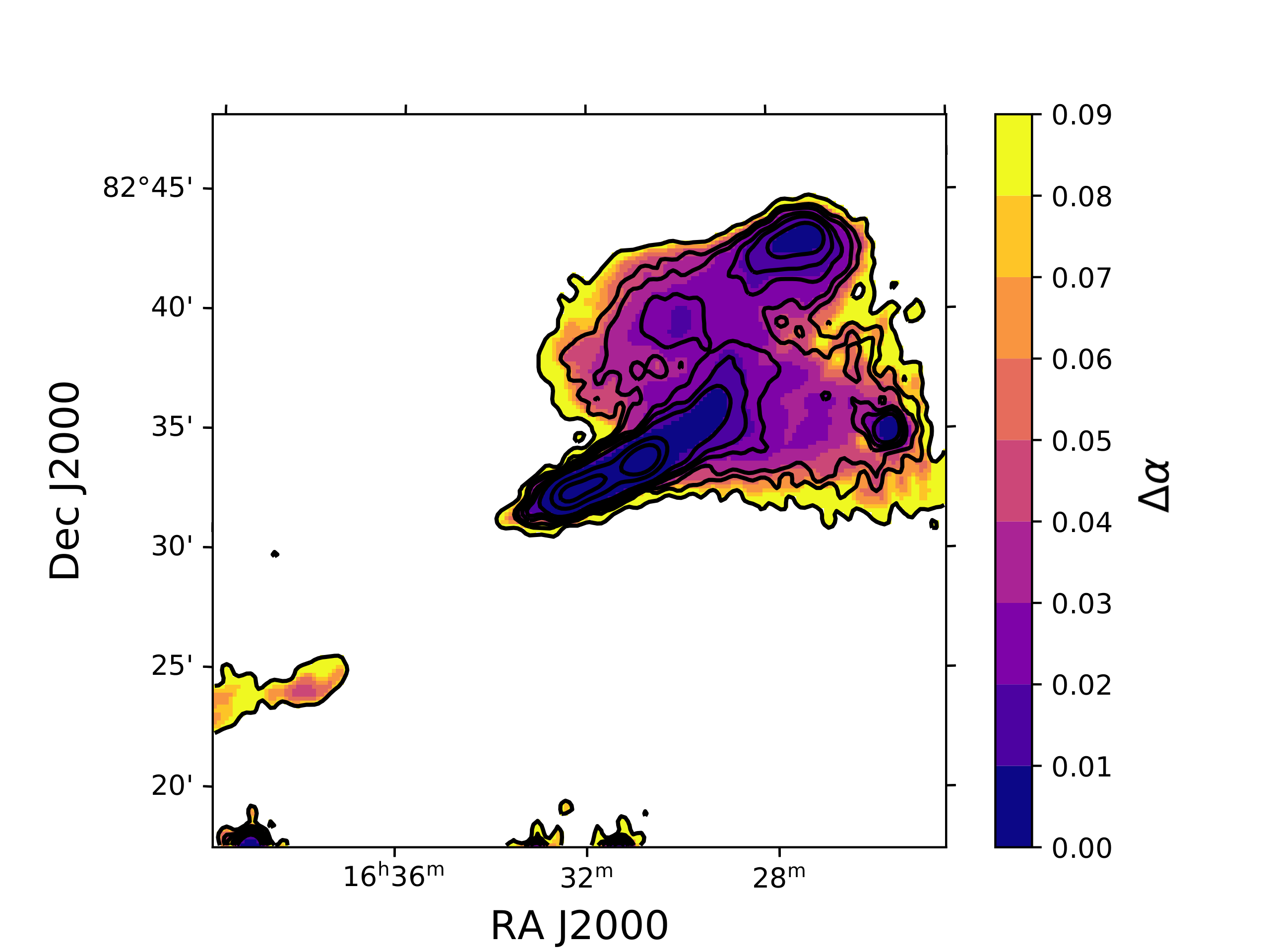}}\\
 \end{tabular}
     \caption{(a) Spectral-index maps between 150\,MHz and 1.4\,GHz for the northern lobe of NGC~6251.  The flux cutoff used was 7$\sigma_{\rm rms}$ where $\sigma_{\rm rms}=1.5$\,mJy\,beam$^{-1}$ is the rms noise of the LOFAR image.  (b) Corresponding spectral-index error map.}
     \label{fig:vla_spix_north}
\end{figure*}

\begin{figure*}
  \centering
  \begin{tabular}{cc}
     \subfloat[][]{\label{fig:VLA_south_spix}}{\includegraphics[width=0.5\textwidth]{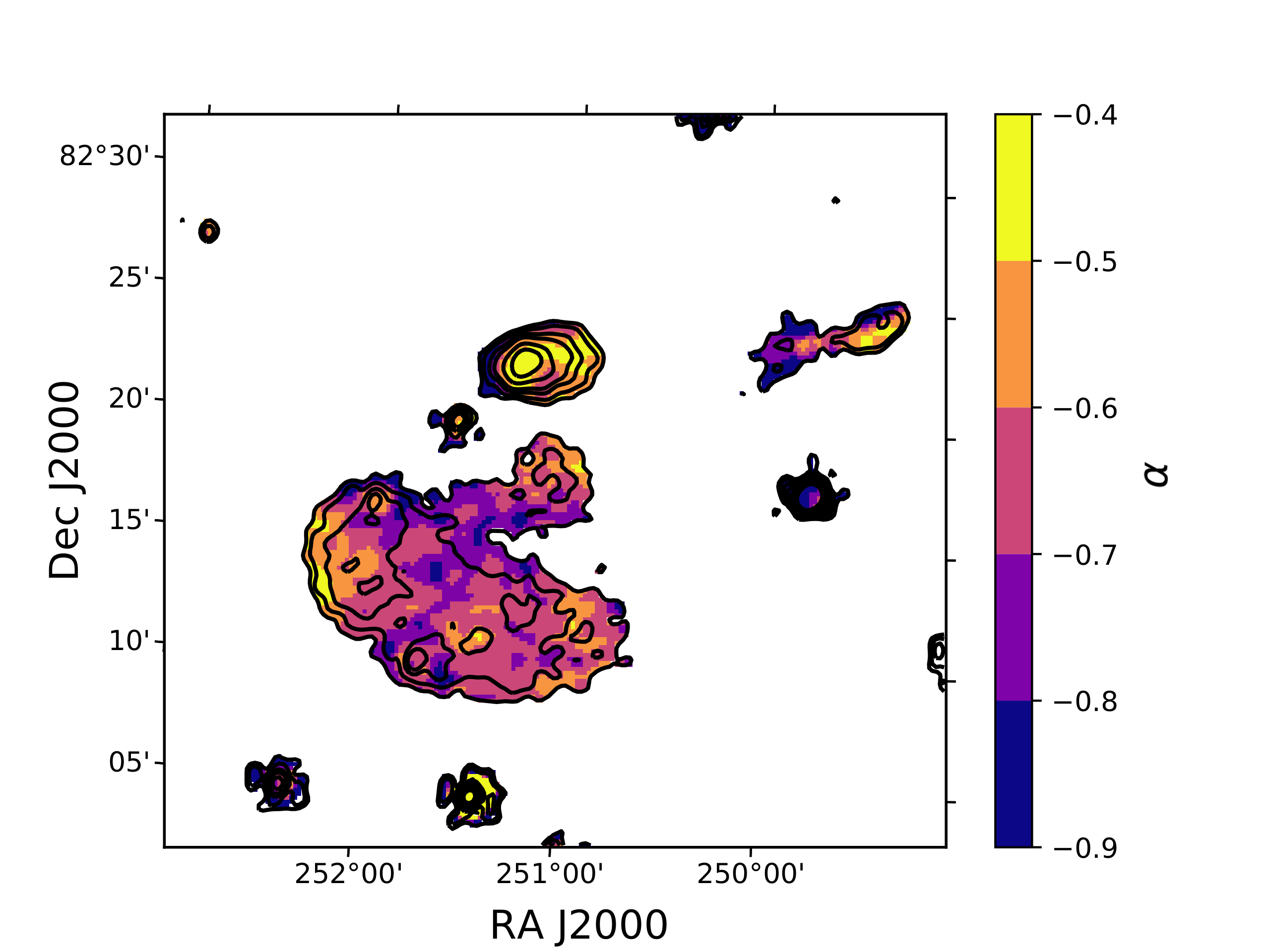}} & \subfloat[][]{\label{fig:VLA_south_spix_error}}{\includegraphics[width=0.5\textwidth]{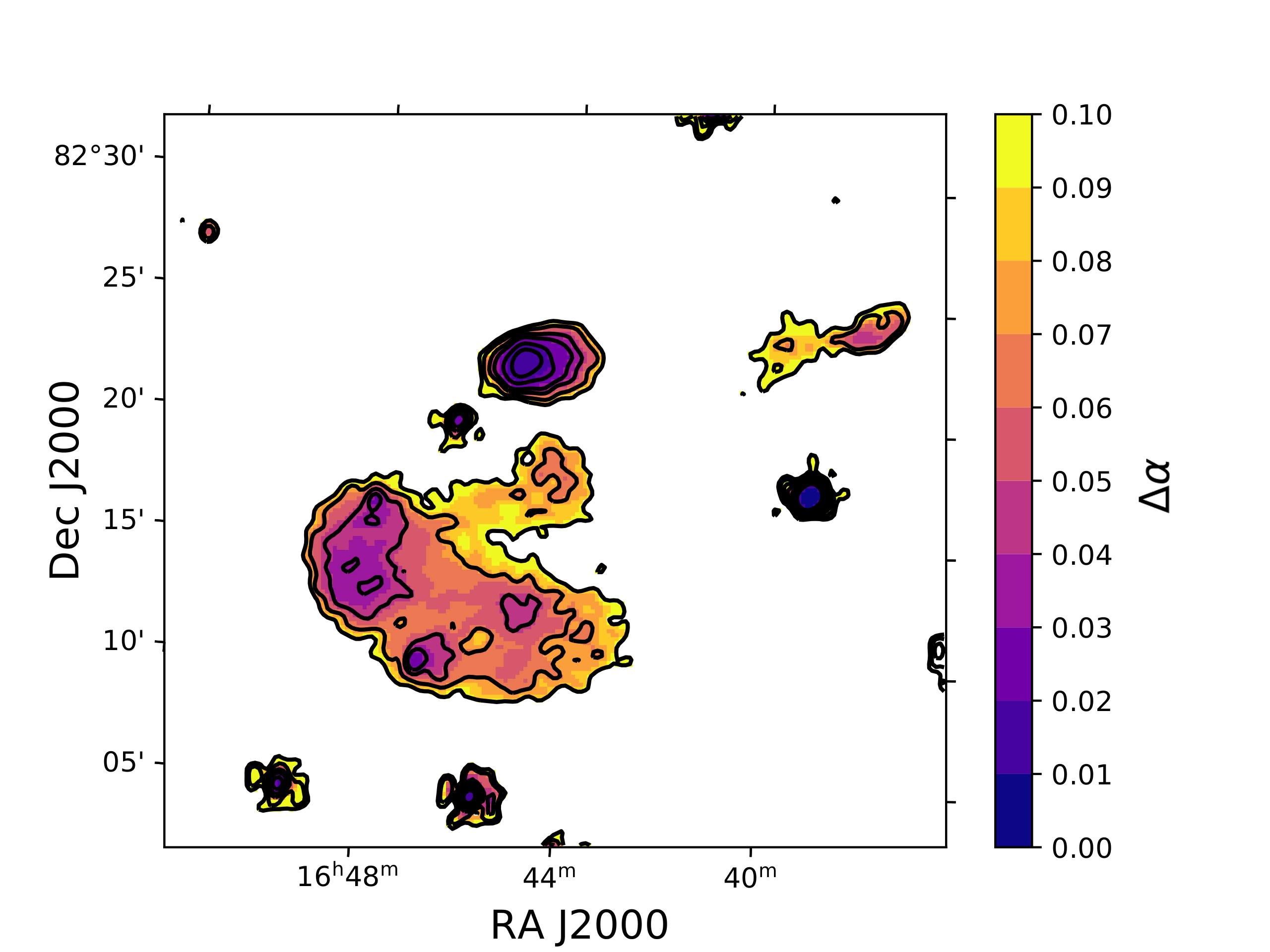}}\\
 \end{tabular}
     \caption{(a) Spectral-index maps between 150\,MHz and 1.4\,GHz for the southern lobe of NGC~6251, with the same flux cutoff as in Fig.~\ref{fig:VLA_north_spix}.  Note that the colour scale differs from Fig.~\ref{fig:VLA_north_spix}.  (b) Corresponding spectral-index error map.}
     \label{fig:vla_spix_south}
\end{figure*}

To investigate the variation of the spectral index across the source and reduce the ambiguity associated with inconsistent \textit{uv}~coverage, the LOFAR data were re-imaged with a \textit{uv}~range of 140--4400 $\lambda$, matching that of the VLA.  This ensures that both total-intensity maps used to calculate the spectral index include emission from the same spatial scales, although they may still differ in the \textit{uv}~coverage of the specific observations.  The resulting LOFAR image has an rms of 1.5\,mJy\,beam$^{-1}$.  A spectral-index map was made from 150\,MHz to 1.4\,GHz using pixels exceeding a 7$\sigma$ limit.  The resulting images are shown in Figs~\ref{fig:vla_spix_north} and~\ref{fig:vla_spix_south}.  Due to the inner \textit{uv}~limit at $140\lambda$, neither the extension nor the southern backflow are visible in the \textit{uv}-matched LOFAR image.

The spectral-index map in Fig.~\ref{fig:vla_spix_north} shows the core of NGC~6251 to have a flat spectrum.  The spectral index is around $-0.5$ along the axis of the inner part of the main jet, steeping on either side.  The jet steepens as it enters the northern lobe to $-0.7$, before flattening to around $-0.5$ in the hotspot.  The spectral index of the northern lobe varies from around $-0.7$ near the jet and hotspot to $<-1$ towards the western extension.

The western extension of the northern lobe is outside the primary beam of the VLA.  The shortest baseline for the WSRT is 36\,m which gives a maximum angular scale of 47\,arcmin at 610\,MHz and 88\,arcmin at 325\,MHz.  The WSRT data should therefore be sensitive to emission on these scales.  The fact that the LOFAR image shows the extension continuing for another 14.4\,arcmin past what is seen in the WSRT 325\,MHz image suggests that the emission has a very steep spectral index, at least steeper than $\alpha=-2.7$.  The emitting electrons are likely very old.

The base of the southern counterjet can also be seen in Fig.~\ref{fig:vla_spix_north}.  The spectral index is $<-0.6$, steeper than in the main jet.  This appears to be the flattest part of the pre-bend region of the counterjet.  Beyond 60\,kpc the counterjet has steepened such that it is only visible at 150\,MHz. 

The counterjet reappears at higher frequencies in what appears to be a bend (see Fig.~\ref{fig:vla_spix_south}).  The spectral index of this bend is around $-0.5$ with a cocoon of steeper emission ($\alpha<-0.9$) surrounding it.  The spectral index for the southern hotspot is almost flat, with $\alpha > -0.5$.  Similar to the extension of the northern lobe, the diffuse low-surface-brightness emission seen around the southern jet in the LOFAR image is not seen in the 325\,MHz WSRT image, suggesting that the spectral index is at least as steep as $-1.6$.  This is steeper than the spectral index seen in the lobe, suggesting that this is ageing material from the lobe being redirected back along the jet axis.

There are substantial discrepancies between the spectral-index map of the southern region of the source in Fig.~\ref{fig:vla_spix_south}(a) and the steeper integrated spectral indices derived from Fig.~\ref{fig:spectral_index_plots}(b) and listed in Table~\ref{tab:flux}.  These may result from inconsistent \textit{uv}~coverage: Fig.~\ref{fig:vla_spix_south}(a) shows spectral indices based only on our \textit{uv}-matched total-intensity maps, whereas the integrated spectral indices also incorporate the maps of \citet{Mack1997}.  They may also result from differences in the assumed location of the emission: the integrated spectral indices are based on the regions defined in Fig.~\ref{fig:NGC6251_regions} including the fringes of the lobe, which fall below the flux cutoff used for the spectral-index maps and might be expected to have systematically older, steeper emission.

\subsection{Polarisation}
\label{sec:pol}

The Stokes~Q/U images produced per Section~\ref{sec:polimg} were analysed with both RM~synthesis (Section~\ref{sec:rmsynth}) and QU~fitting (Section~\ref{sec:QUfitting}).

\subsubsection{Rotation-measure synthesis}
\label{sec:rmsynth}

RM~synthesis \citep{Brentjens2005} involves direct calculation of the Faraday spectrum from Stokes~Q/U data.  It is computationally efficient, and not dependent on specific model assumptions, but the output can be misleading when there are multiple Faraday structures along the line of sight \citep{Farnsworth2011}, and it is subject to limitations in scale and resolution for any realistic finite bandwidth.  The maximum observable Faraday depth, $\phi_{\rm max-depth}$, the resolution in Faraday space, $\delta \phi$, and the largest scale in Faraday space that can be detected, $\phi_{\rm max-scale}$, are \citep{Brentjens2005}
\begin{subequations}
  \begin{align}
    \| \phi_{\rm max-depth} \| &\approx \frac{\sqrt{3}}{\delta \lambda^{2}}\\
    \delta \phi &\approx \frac{2\sqrt{3}}{\Delta \lambda^{2}}\\
    \phi_{\rm max-scale} &\approx \frac{\pi}{\lambda_{\rm min}^{2}},\label{max-scale}
  \end{align}
\end{subequations}
where $\delta \lambda^{2}$ is the width of a channel in $\lambda^2$, $\Delta \lambda^{2}$ is the total width of the $\lambda^{2}$ coverage and $\lambda_{\rm min}^{2}$ is the minimum value of $\lambda^2$.  For our observations this gives $|\phi_{\rm max-depth}| \approx 677$\,rad\,m$^{-2}$, $\delta \phi \approx 0.87$\,rad\,m$^{-2}$ and $\phi_{\rm max-scale} \approx 0.46$\,rad\,m$^{-2}$.

\begin{figure}
 \centering
 \includegraphics[width=0.5\textwidth]{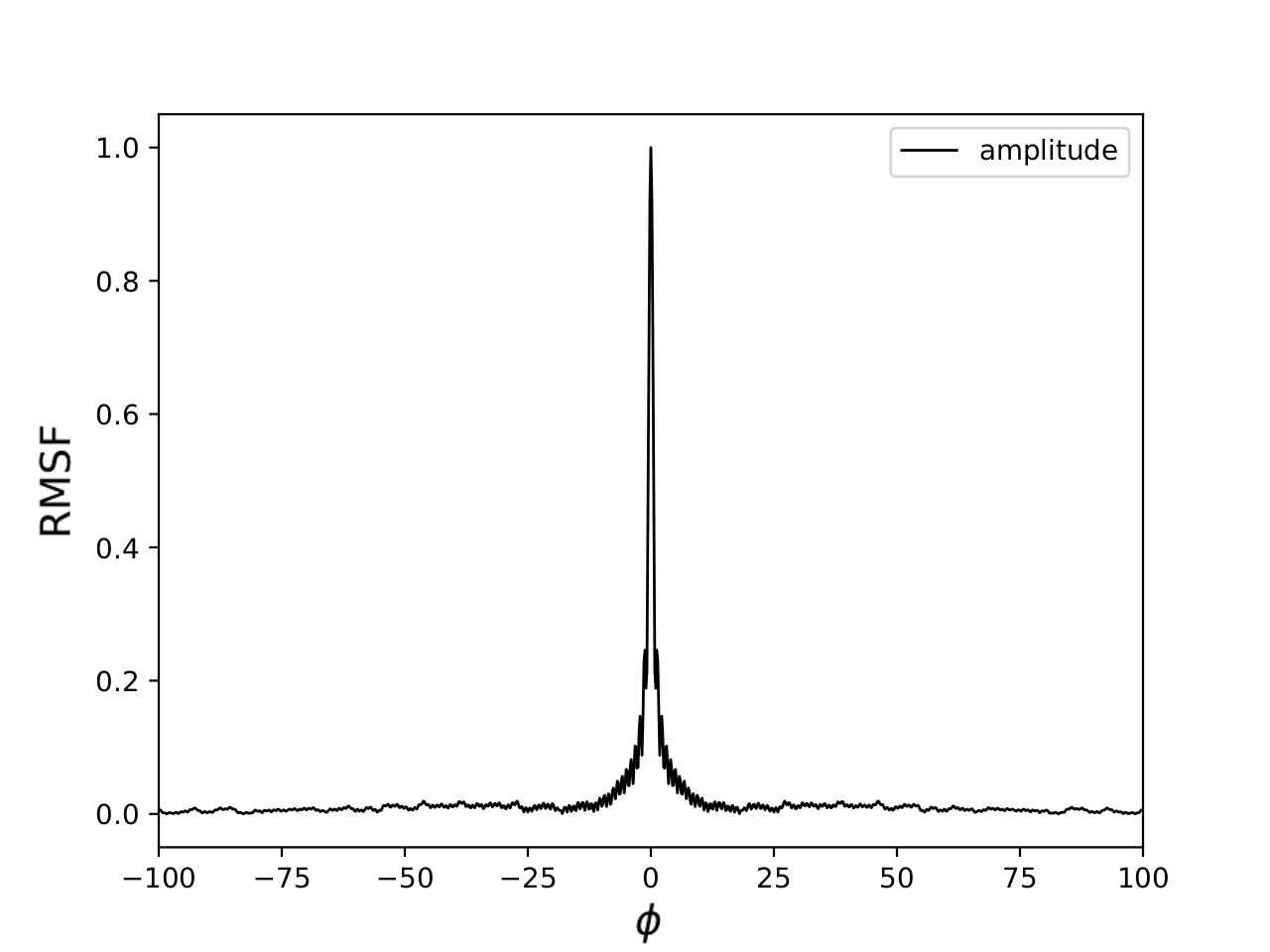}
  \caption{RMSF for the LOFAR HBA data as a function of Faraday depth $\phi$ (rad\,m$^{-2}$).}
  \label{fig:rmsf}
\end{figure}

We calculated the Faraday spectrum from the Stokes~Q and~U images, neglecting spectral dependence of the polarised flux, using the RM~synthesis code \textsc{pyrmsynth}\footnote{\url{https://github.com/mrbell/pyrmsynth}}.  In our first iteration, we searched the entire range of Faraday depths to which our observations were sensitive, from $-1000$\,rad\,m$^{-2}$ to $+1000$\,rad\,m$^{-2}$, using a coarse Faraday-depth cell size of 2\,rad\,m$^{-2}$.  From this spectrum we excluded any structure at large Faraday depths.  In our second iteration, we searched over Faraday depths from $-300$\,rad\,m$^{-2}$ to $+300$\,rad\,m$^{-2}$ with a cell size of 0.2\,rad\,m$^{-2}$ to properly sample the rotation-measure spread function (RMSF).  Fig.~\ref{fig:rmsf} shows the RMSF of the LOFAR data.

Following \citet{VanEck2018}, we fit a Rayleigh distribution to the Faraday spectrum of each pixel in order to estimate the noise in the Faraday spectra.  Faraday depths between $-20$\,rad\,m$^{-2}$ and $+20$\,rad\,m$^{-2}$ were masked to avoid fitting the instrumental polarisation.  The scale parameter $\sigma$ was assumed to be the noise in the spectrum.  An $8\sigma$ detection threshold was applied to the spectrum of each pixel.

\begin{figure}
  \centering
  \begin{tabular}{c}
     \subfloat{\label{fig:faraday_spect_src}}{\includegraphics[width=0.5\textwidth]{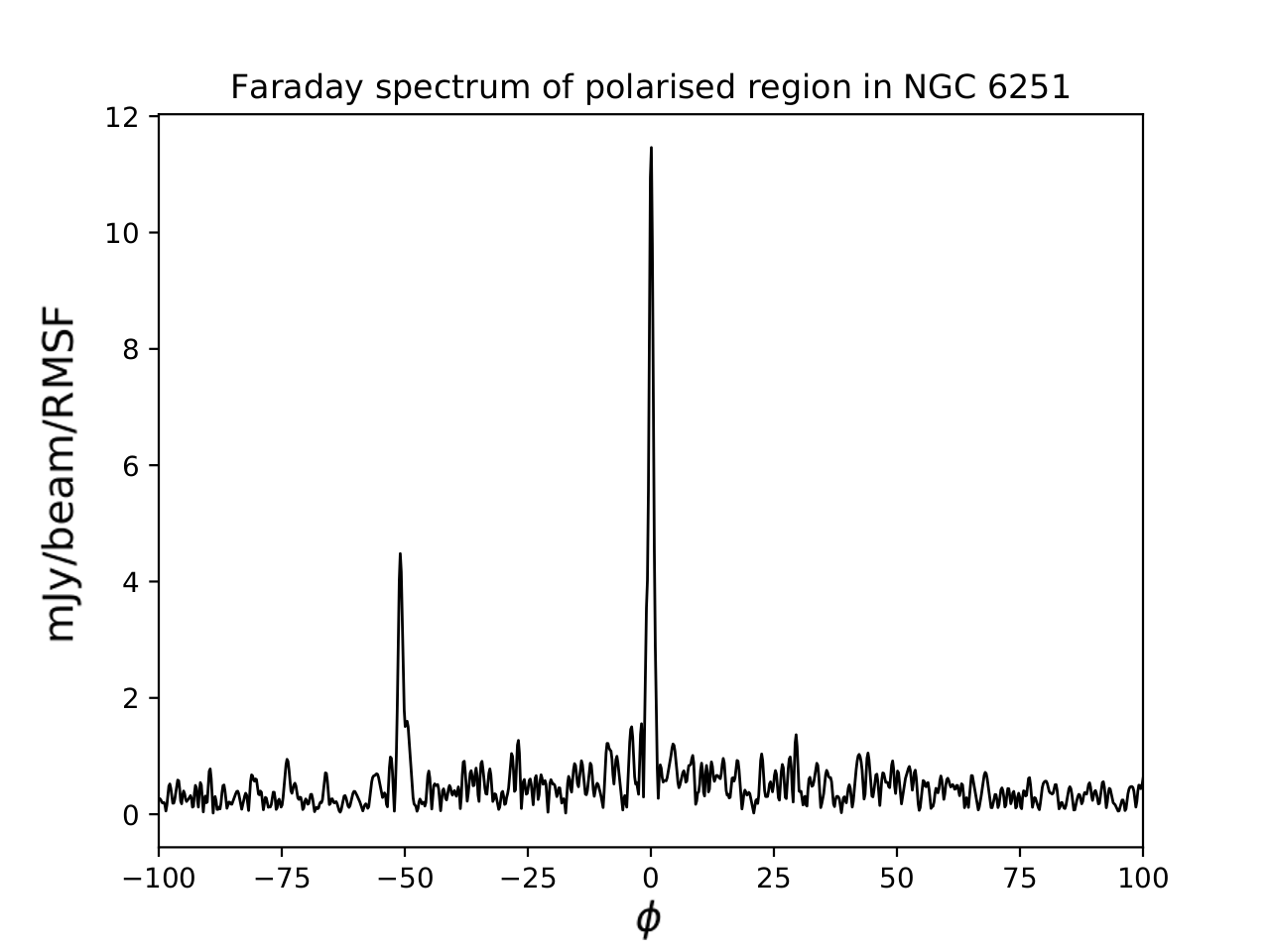}}\\
     \subfloat{\label{fig:faraday_spect_non}}{\includegraphics[width=0.5\textwidth]{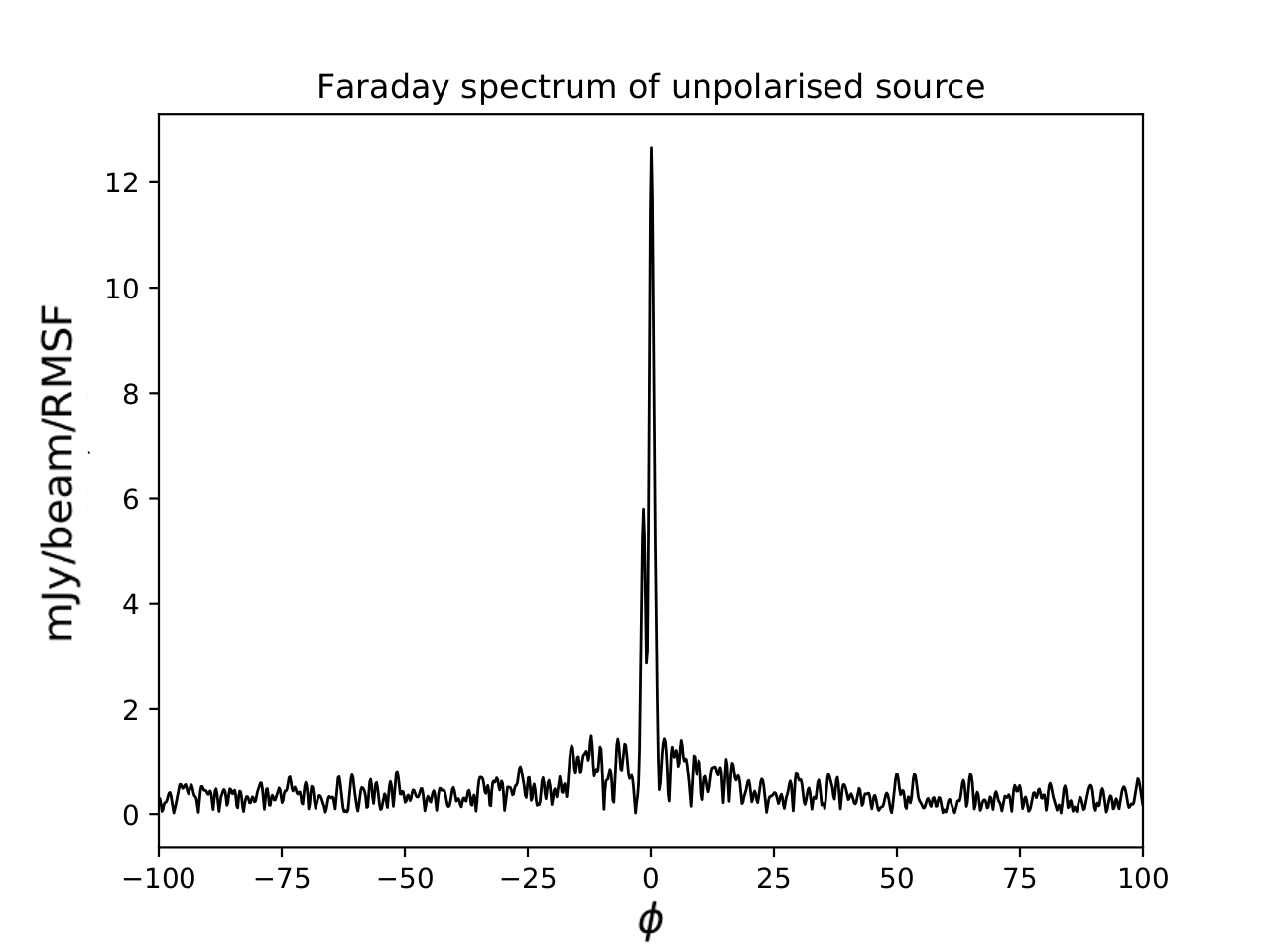}} \\
 \end{tabular}
     \caption{Faraday spectra as a function of Faraday depth $\phi$ (rad\,m$^{-2}$) for (a) a polarised region in NGC~6251; and (b) an unpolarised source.  In both cases there is a strong peak centred on $\phi \sim 0$\,rad\,m$^{-2}$ which corresponds to unpolarised emission misidentified as polarised due to instrumental polarisation.  In the first case, there is also a peak at $\phi \sim 50$\,rad\,m$^{-2}$ which represents polarised, Faraday-rotated emission from NGC~6251.}
     \label{fig:full_FS}
\end{figure}

\begin{figure}
 \centering
 \includegraphics[width=0.5\textwidth]{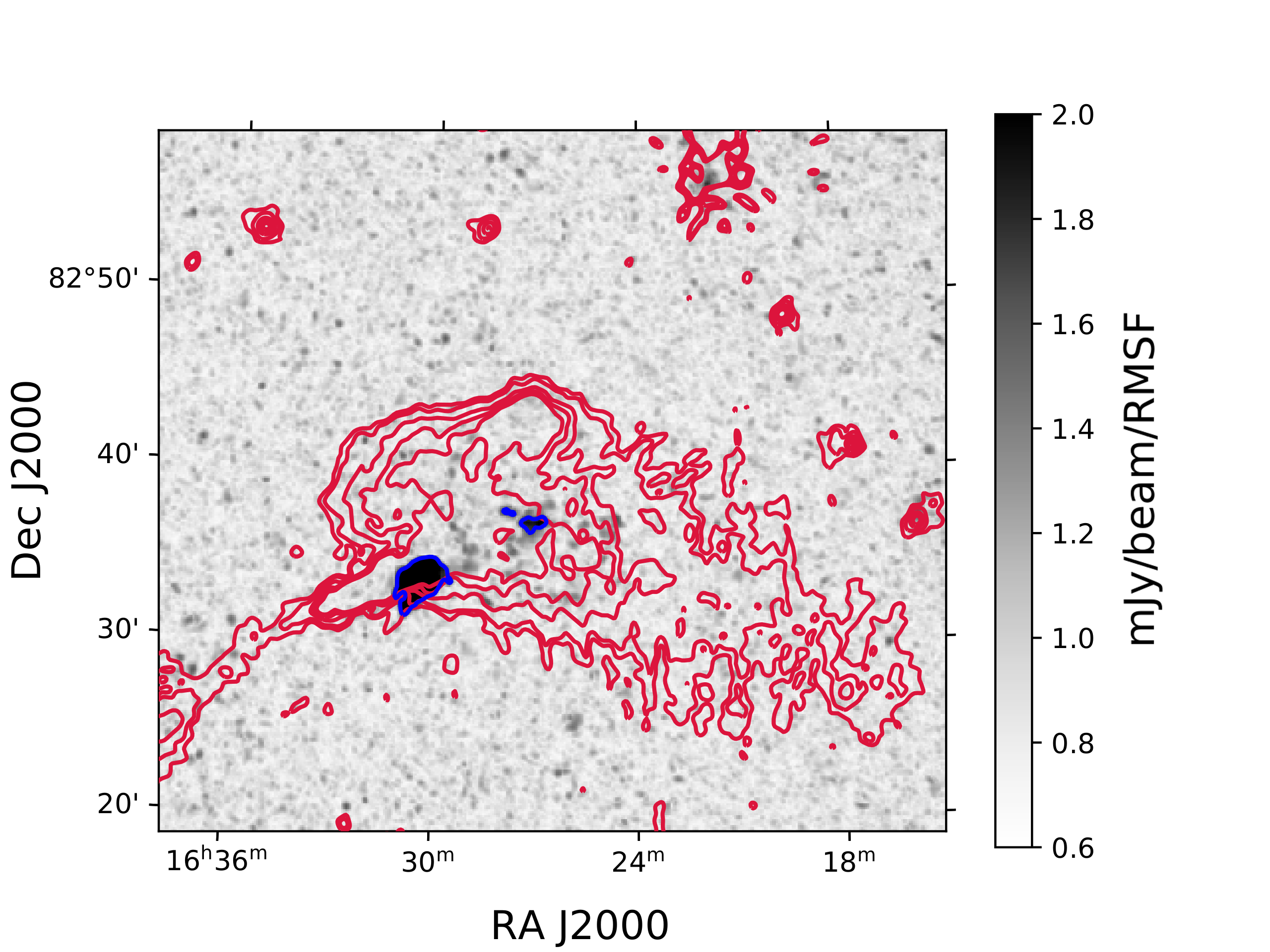}
 \caption{Polarisation in the northern lobe of NGC~6251.  Greyscale shows peak polarisation in the Faraday depth cube, with blue contours at 8$\sigma$.  Red contours show unpolarised emission in the same LOFAR data at 3, 5, 10, 15, 20~$\times$~$\sigma_{\rm rms}$ where $\sigma_{\rm rms}=2.0$\,mJy\,beam$^{-1}$.  Faraday spectra for representative pixels in the polarisation-detected knot and lobe are shown in Fig.~\ref{fig:FS}.}
 \label{fig:max_pol}
\end{figure}

Fig.~\ref{fig:full_FS} illustrates typical Faraday spectra from both an unpolarised field source and a polarised region of NGC~6251.  Faraday spectra like these, calculated per Section~\ref{sec:pol} for each pixel in the image, constitute a Faraday cube.  Fig.~\ref{fig:max_pol} shows the polarised intensity at the maximum in the Faraday spectrum for each pixel, after excluding the region $-15$\,rad\,m$^{-2} < \phi < +15$\,rad\,m$^{-2}$ to exclude the instrumental polarisation.  The blue contours mark the regions where the peak in the Faraday spectrum is $>8\sigma_{\rm rms}$, where $\sigma_{\rm rms}$ is the noise in the Faraday spectrum of that pixel.  There is a clear detection of polarisation in the knot of the jet, with peak intensity at 16h30.5m 82$^\circ$33.3', as well as some patchy structure in the northern lobe.  The rest of the source is depolarised.

\begin{figure}
  \centering
  \begin{tabular}{c}
     \subfloat[][]{\label{fig:knot_fs}}{\includegraphics[width=0.5\textwidth]{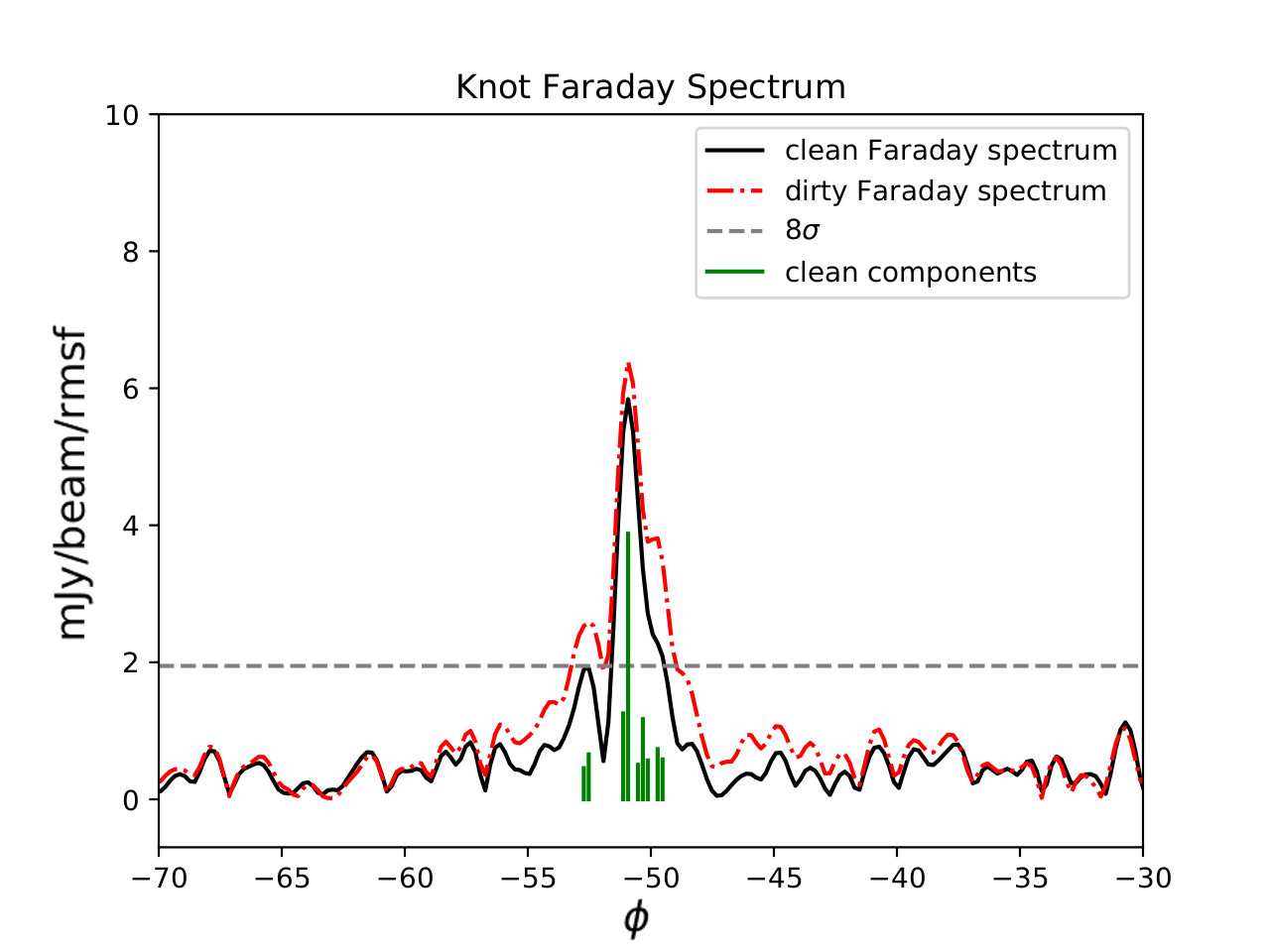}} \\
     \subfloat[][]{\label{fig:lobefs}}{\includegraphics[width=0.5\textwidth]{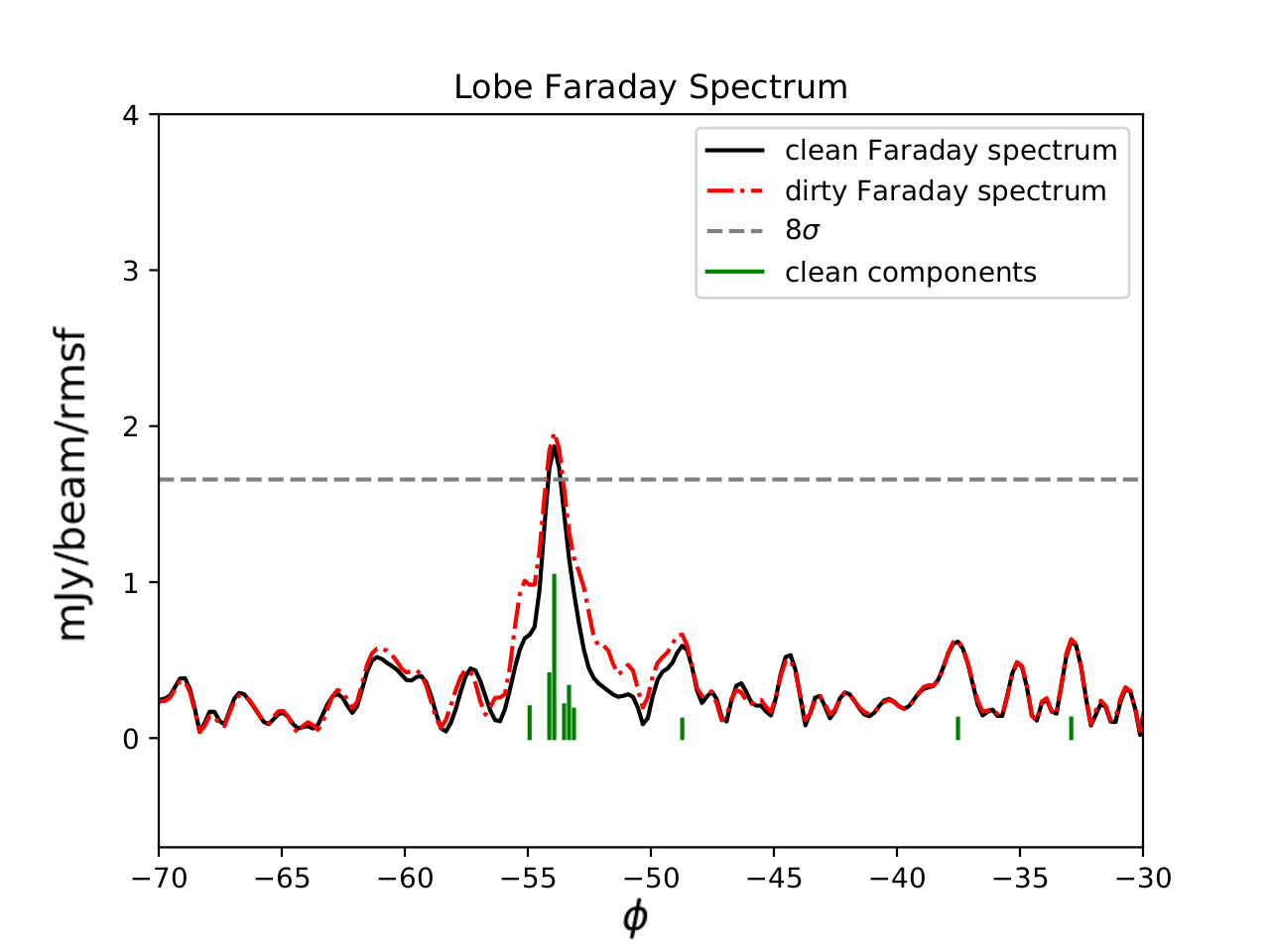}}\\
 \end{tabular}
     \caption{Faraday spectra for pixels in the polarisation-detected (a) knot and (b) lobe in Fig.~\ref{fig:max_pol}.  The dirty Faraday spectra are shown as in Fig.~\ref{fig:full_FS}; also shown are the spectra after cleaning.}
     \label{fig:FS}
\end{figure}

Fig.~\ref{fig:FS} shows the Faraday spectrum for a representative pixel in the knot and a representative pixel in the lobe.  The Faraday spectrum of the lobe shows a single Faraday-thin component.  The average Faraday depth of this component for the pixels in which a polarisation detection has been made is $-54.1$\,rad\,m$^{-2}$ with a standard deviation of $0.4$\,rad\,m$^{-2}$.  The mean amplitude of the Faraday-thin component in the lobe is $1.8 \pm 0.1$\,mJy\,beam$^{-1}$\,RMSF$^{-1}$.  

The knot shows a single Faraday-thin component with an average Faraday depth of $-50.97$\,rad\,m$^{-2}$ and a standard deviation of 0.07\,rad\,m$^{-2}$.  The average amplitude of this component is $3$\,mJy\,beam$^{-1}$\,RMSF$^{-1}$ with a standard deviation of $1$\,mJy\,beam$^{-1}$\,RMSF$^{-1}$.  The fractional polarisation is approximately 1\%.  \citet{Perley1984} report typical polarisation fractions in the jet of $\sim 10$\% at 1662\,MHz, rising to a peak $\sim 40$\% around the position of the knot (their Fig.~15).  Our lower polarisation fraction may result from a combination of beam depolarisation due to variation in the polarisation angle across the jet (their Fig.~17), and our inclusion of unpolarised emission from the steeper-spectrum emission on either side of the jet, due both to our lower frequency and our larger beam size.

\citet{Perley1984} also report polarised emission at 1662\,MHz with a polarisation fraction $\sim 20$\% from the inner jet, which does not appear in Fig.~\ref{fig:max_pol}.  This may be caused by Faraday depolarisation, either inherent to the source or resulting from the group environment, which would suppress the polarised signal at our lower frequency of 150\,MHz.  Note that \citeauthor{Perley1984} measure strong RM gradients in this region (see their Fig.~20a).

\subsubsection{QU fitting}
\label{sec:QUfitting}

QU fitting involves fitting parameters of a modelled polarised source to reproduce observed Stokes~Q/U data.  A model for the polarised intensity $P(\lambda^{2}) = Q+iU$ of a Faraday-thin source can be expressed as
\begin{equation} \label{eqn:far_thin}
  P(\lambda^{2})=p_0\exp\left[2i\left(\chi+\phi \lambda^{2}\right)\right],
\end{equation}
where $\chi$ is the polarisation angle and $p_0$ the initial polarised intensity.  More complex models can be constructed e.g.\ from the superposition of multiple such sources.  QU fitting typically achieves more precise reconstruction of source parameters (including Faraday depth) than RM~synthesis \citep{sun2015}, but is dependent on selection of the correct model, and is more computationally intensive.  Here we use QU fitting as a follow up, to further investigate the structure of the polarised emission described in Section~\ref{sec:rmsynth}, and to more precisely reconstruct its Faraday depth.  We use the \textsc{qu-jb} code presented by \citet{sun2015}, which explores the parameter space and evaluates Bayesian evidence using the \textsc{MultiNest} library \citep{Feroz2008,Feroz2009,Feroz2013}.

We fitted a number of different models to our data.  First, our null hypothesis was that only instrumental polarisation is present, which we modelled as a 1st-order polynomial in frequency space.  This order was chosen as it was sufficient to adequately suppress the instrumental polarisation, while insufficient to fit out or degrade an astronomical signal at significantly non-zero Faraday depth.  We then fitted a series of models with both this instrumental polarisation and one, two or three Faraday-thin components, each with a contribution to the polarised emission as given in equation~(\ref{eqn:far_thin}), approximated to be independent of frequency over our band.  We did not fit for any models with Faraday-thick components, as LOFAR HBA is minimally sensitive to these: from equation~(\ref{max-scale}) the maximum scale recoverable is $\phi_{\rm max-scale}=0.46$\,rad\,m$^{-2}$, so any polarised source with Faraday thickness exceeding this would not be properly recovered, although the edges in its Faraday spectrum might be visible if they were sufficiently sharp \citep{vanEck2017}.

In order to evaluate the quality of the fits across the knot and the lobe we calculate for each pair of models the Bayes factor, $K$, which is given by
\begin{equation}
K = \frac{\textrm{Pr}\left(D|M_{1}\right)}{\textrm{Pr}(D|M_{2})} = \frac{\int \textrm{Pr}(\theta_{1}|M_{1}) \, \textrm{Pr}(D|\theta_{1},M_{1}) \, d\theta_{1}}{\int\textrm{Pr}(\theta_{2}|M_{2}) \, \textrm{Pr}(D|\theta_{2},M_{2}) \, d\theta_{2}},
\end{equation}
where each model~$M_i$ is defined in terms of parameters $\theta_i$, is assigned a prior Pr$(\theta_{i}|M_{i})$, fits the data with likelihood Pr$(D|\theta_{i},M_{i})$, and is supported by Bayesian evidence Pr$\left(D|M_{i}\right)$.  Following \citet{Kass90}, we evaluate the Bayes factor based on the derived value 2ln$K$: a positive value supports model~2, constituting weak (values $<2$), positive (2--6), strong (6--10) or very strong ($>$10) evidence.  Negative values of 2ln$K$, similarly, support model~1.

\begin{figure*}
  \centering
  \begin{tabular}{cc}
     \subfloat[][]{\label{fig:p0n1}}{\includegraphics[width=0.5\textwidth]{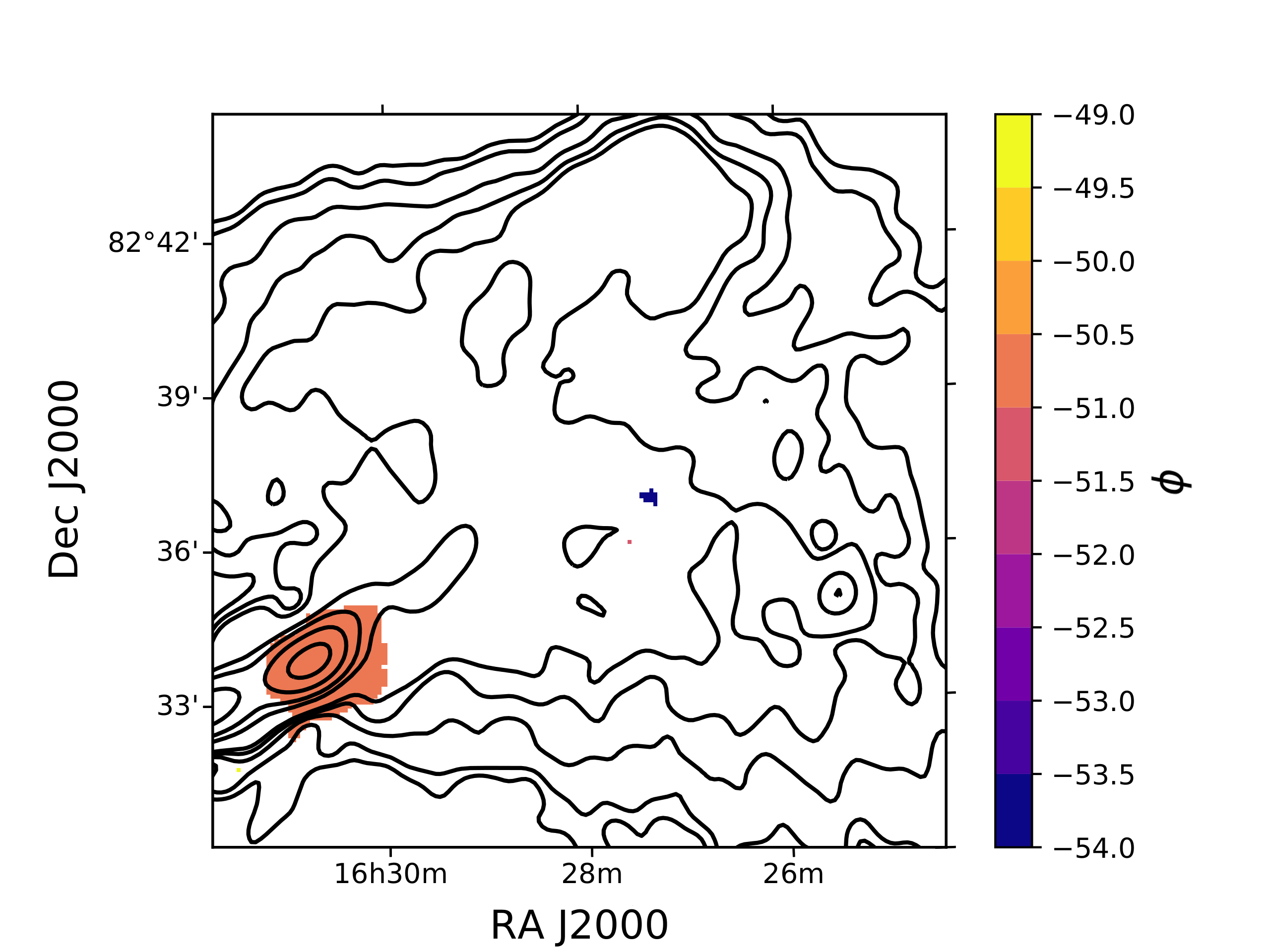}} & \subfloat[][]{\label{fig:bayesp0n1}}{\includegraphics[width=0.5\textwidth]{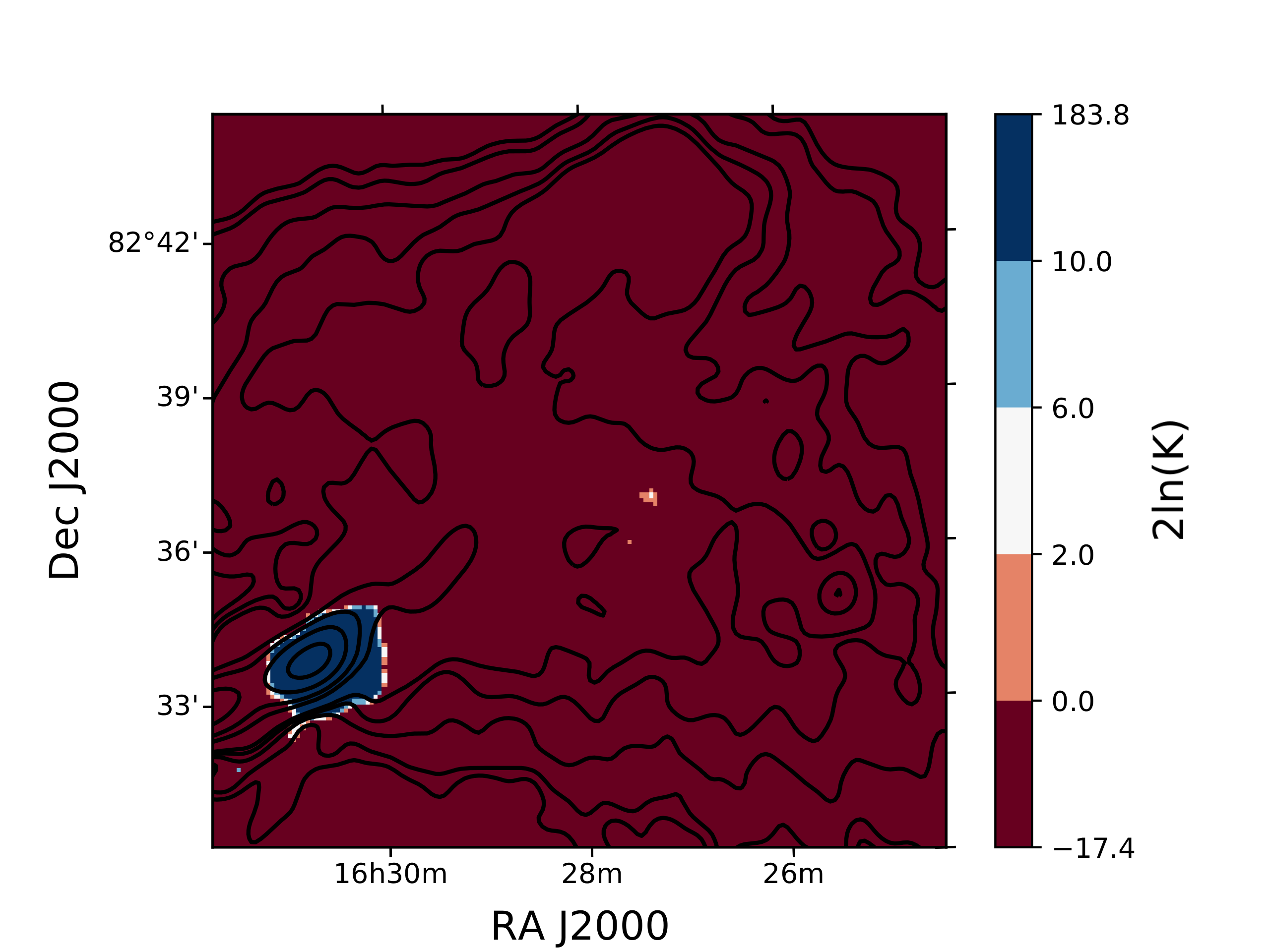}}\\
 \end{tabular}
     \caption{Results of the QU fitting.  (a) shows the Faraday depth of the component found for a single Faraday-thin screen plus instrumental polarisation.  (b) The Bayes factor when comparing the fits for a single thin screen plus instrumental polarisation with just instrumental polarisation.  Red indicates support for the null hypothesis while blue indicates support for the single Faraday-thin screen.  White indicates inconclusive.}
     \label{fig:QU_fits}
\end{figure*}

Fig.~\ref{fig:QU_fits} shows the results for the model with one Faraday-thin source.  The evidence of the models with two or three Faraday-thin sources is less than the evidence for all other tested models and so is not shown.  Fig.~\ref{fig:p0n1} shows the Faraday depth found for the one-source model for each pixel where it is favoured over the null hypothesis with $K>1$.  Fig.~\ref{fig:bayesp0n1} shows $2\textrm{ln}K>2$, where white to blue ($2\textrm{ln}K>2$) indicates support for the one-source model and red ($2\textrm{ln}K<2$) indicates support for the null hypothesis.  We find an average Faraday depth in the knot of $-50.8$\,rad\,m$^{-2}$.  Despite the fact that there is a significant polarised region detected at $5\sigma$ in the Faraday spectrum of the lobe, only 10~pixels are detected in the lobe with the presence of a source favoured over the null hypothesis with $K>1$.  The average Faraday depth of these pixels is $-53.6$\,rad\,m$^{-2}$, in agreement with that measured from the Faraday spectra, with a standard deviation of 0.7\,rad\,m$^{-2}$.  The difference of 2.8\,rad\,m$^{-2}$ between the Faraday depths in the knot and the lobe is significant against this uncertainty.

\begin{figure}
 \centering
 \includegraphics[width=0.5\textwidth]{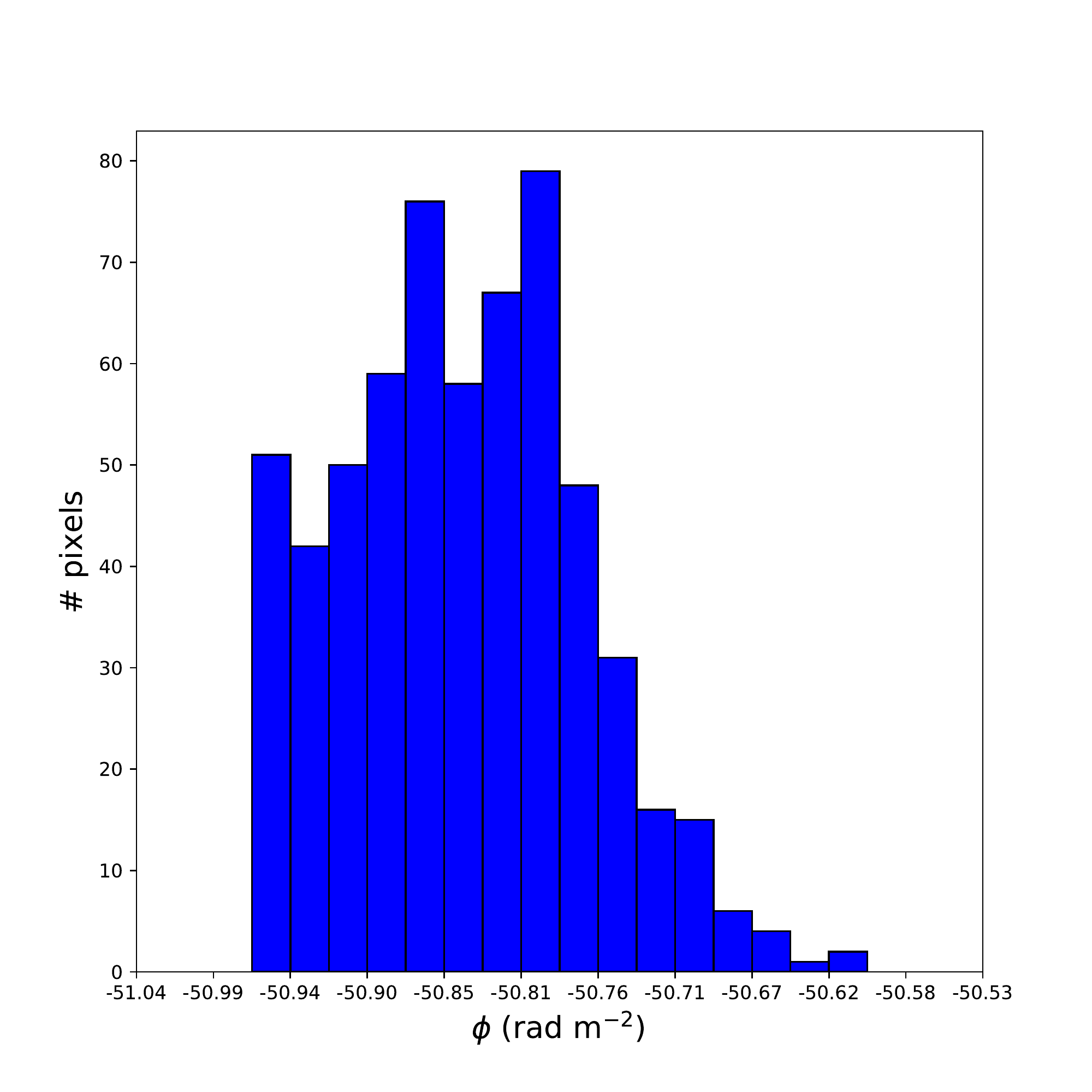}
  \caption{Distribution of Faraday depths in the knot in main jet of NGC~6251.}
  \label{fig:RM_distribution}
\end{figure}

There are small variations in Faraday depth of order $\Delta\phi\sim0.2$\,rad\,m$^{-2}$.  Fig.~\ref{fig:RM_distribution} shows the distribution of the Faraday depths in the knot.  The variance in the Faraday depth is $\sigma_{\rm RM}^2=5\times10^{-3}$\,rad$^{2}$\,m$^{-4}$.  If the variation in Faraday depth is due solely to noise/measurement error then the expected variance can be calculated as
\begin{equation}
	\sigma_{\rm RM,noise}^2=\frac{\sum{\sigma_{\rm \phi,i}^{2}}}{N_{\rm beam}}
\end{equation}
where $\sigma_{\phi_{\rm i}}$ is the error in Faraday depth for pixel $i$ and $N_{\rm beam}$ is the number of beams covering the region.  We find that the expected variance is $\sigma_{\rm RM,noise}=2\times10^{-2}$\,rad$^{2}$\,m$^{-4}$.  That $\sigma_{\rm RM,noise}^2$ is so much larger than $\sigma_{\rm RM}^2$ shows that the measurement errors are being overestimated.  This is to be expected as we are unable to properly account for the instrumental polarisation in the Q and U data.  Our inability to accurately model the instrumental polarisation as a simple polynomial is most likely due to inaccuracy in the ionospheric calibration with \textsc{RMExtract}.  This has the effect of shifting the instrumental polarisation away from zero by different amounts as a function of time \citep{VanEck2018}.  This leads to a wide posterior distribution.  However, for the structure to be real, the estimated errors would need to be 3~times larger than the actual uncertainty in Faraday depth.

\section{Discussion and analysis}
\label{sec:discussion}

\subsection{Spectral index and energetics}
\label{sec:disc_specind}

The spectral-index maps of NGC~6251 presented in Section~\ref{sec:alpha} are the first to extend down to 150\,MHz at this angular resolution, permitting detailed modelling of the electron populations responsible for this emission.  In this Section, we perform this modelling to determine the pressure, age and energetics of electrons in the lobes of NGC~6251, and discuss these results with reference to the literature.

As a check, we compare our 150\,MHz--1.4\,GHz spectral-index maps presented here to the 325--610\,MHz and 408\,MHz--10\,GHz maps of \citet{Mack1998}.  We find that our map agrees well with their 325--610\,MHz map except for the bend in the southern jet.  Here we find a spectral index of $\alpha \sim -0.5$, whereas \citeauthor{Mack1998} find $\alpha < -1$.  The 150\,MHz LOFAR image shown in Fig.~\ref{fig:HBA_total_Intensity} shows the backflow of the lobe material from the southern lobe.  In the 325\,MHz WSRT image only a small region of this structure is detected and in the 610\,MHz image only the bend is visible.  We suggest that the presence of the older lobe emission, coincident with the jet, has led to the steep spectral index in the \citeauthor{Mack1998} spectral-index map and that the bend is indeed a real feature of the counterjet.

\subsubsection{Internal pressure and magnetic field}
\label{sec:pressure}

The internal pressure of a relativistic plasma can be calculated from its energy density, with contributions from relativistic electrons ($U_{\rm e}$), protons ($U_{\rm p}$) and the magnetic field ($U_B$).  Assuming equipartition between the magnetic field and the relativistic particles, and defining the ratio \mbox{$k = U_{\rm p} / U_{\rm e}$}, which we take to be constant, then the internal pressure is
 \begin{equation} \label{eq:Pint}
  P_{\rm int}=\frac{k+2}{3}U_B .
 \end{equation}
We calculate $U_{\rm e}$ and $U_B$ using the \textsc{synch} code \citep{Hardcastle1998}, with the spectral indices from Section~\ref{sec:alpha} as inputs.  Briefly, this code calculates the energetics of a relativistic plasma in equipartition given a measurement of the radio flux, the proton/electron energy-density ratio $k$, and a power-law model of the electron energy distribution, including minimum and maximum energies of the population and, optionally, a spectral break.  From these, it calculates the equipartition magnetic field strength, and hence the energy density of the magnetic field and electron population.

We applied \textsc{synch} to find $U_B$ under these assumptions for each component of NGC~6251.  We assumed protons to be absent ($k=0$) and took the low-energy and high-energy cutoffs for the electron population to be, respectively, $5\times 10^{6}$\,eV and $5\times10^{11}$\,eV.  The low-energy value corresponds to a minimum Lorentz factor, below which synchrotron losses are unimportant, of \mbox{$\gamma_{\rm min} = 10$}.  Investigations of hotspots suggest values of $\gamma_{\rm min} \sim 10^2$ \citep{Barai2006}, or typical values around $10^2$ with occasional values up to $10^4$, but we expect the minimum Lorentz factors in lobes to be lower than in hotspots due to adiabatic expansion.  Within this energy range, we assumed an injection index of $p=-0.6$, consistent with the synchrotron spectral index $\alpha = (p-1)/2 \sim -0.8$ in the hotspots and main jet, with a break energy at which the spectrum steepens to the observed spectral index for other components.  With the energy range fixed, we found the available radio data were best fit with a break energy of $1\times10^9$\,eV.  Finally, we assumed a spherical, elliptical or cylindrical volume for each component as seemed appropriate based on the LOFAR image.

Our assumptions and the resulting fitted pressure for each component are shown in Table~\ref{tab:pressure}.  As the northern extension and southern backflow are likely populated by a very old population of electrons, with limited spectral information available, these data were fitted assuming high-energy cutoffs chosen for consistency with the 325\,MHz radio limit, which were $1\times10^{10}$\,eV for the backflow and $2\times10^{9}$\,eV for the extension.

\begin{table*}
  \centering
  \begin{threeparttable}[b]
  \caption{Parameters for the calculation and evaluation of the internal pressure $P_{\rm int}$ for each component of NGC~6251 (see Section~\ref{sec:pressure}).  Each component is defined in Fig.~\ref{fig:NGC6251_regions}, and its apparent shape in this image is used to calculate its volume, from which we determine the equipartition magnetic field $B_{\rm eq}$ and internal pressure $P_{\rm int}$.  The external pressure $P_{\rm ext}$ is as measured at the projected distance of each component; the ratio $P_{\rm ext}/P_{\rm int}$ indicates whether the component is under- or overpressured.}
  \label{tab:pressure}
  \begin{tabular}{@{}rllllll@{}}
    \toprule
    Component & Shape & Volume (m$^{3}$) & $B_{\rm eq}$ ($\mu$G) & $P_{\rm int}$ (Pa) & $P_{\rm ext}$  (Pa) & $\frac{P_{\rm ext}}{P_{\rm int}}$\\
    \midrule
Core Region        		            & --- & --- & --- & --- & --- & ---\\
Inner Jet (core subtracted)         & Cylindrical & $1.07\times10^{64}$ & 2.5 & $1.7\times10^{-14}$& $4.8\times10^{-13}$ & 28 \\
Knot                                & Cylindrical & $4.88\times10^{63}$ & 3.5 & $3.3\times10^{-14}$& $1.2\times10^{-13}$ & \phn4\\ 
Outer Jet                           & Cylindrical & $9.84\times10^{63}$ & 2.7 & $1.9\times10^{-14}$& $2.4\times10^{-14}$ & \phn1.2\\ 
Northern Lobe                       & Spherical   & $1.55\times10^{66}$ & 1.2 & $3.7\times10^{-15}$& $7.9\times10^{-15}$ & \phn2\\
Northern Extension                  & Cylindrical & $9.1\times10^{65}$ & 1.4 & $4.9\times10^{-15}$ & $4.9\times10^{-16}$ & \phn0.1\\
Northern Hotspot                    & Ellipsoid  & $5.43\times10^{63}$ & 3.2 & $2.7\times10^{-14}$& $3.4\times10^{-15}$ & \phn0.1\\
Southern Jet                        & Cylindrical & $3.24\times10^{64}$ & 1.0 & $3.4\times10^{-15}$& $5.2\times10^{-14}$ & \phn15\\
Southern Backflow                   & Cylindrical & $1.29\times10^{65}$ & 0.8 & $1.6\times10^{-15}$& $2.8\times10^{-15}$ & \phn1.7\\
Southern Lobe                       & Spherical   & $1.32\times10^{66}$ & 1.3 & $4.8\times10^{-15}$& $4.9\times10^{-16}$ & \phn0.1\\
Southern Hotspot                    & Ellipsoid  & $7.09\times10^{63}$ & 2.6 & $1.8\times10^{-14}$& $9.5\times10^{-16}$ & \phn0.05\\
   \bottomrule
 \end{tabular}
\end{threeparttable}
 \end{table*}

\begin{figure*}
  \centering
  \begin{tabular}{cc}
     \subfloat[][]{\label{fig:pressure_north}}{\includegraphics[width=0.5\textwidth]{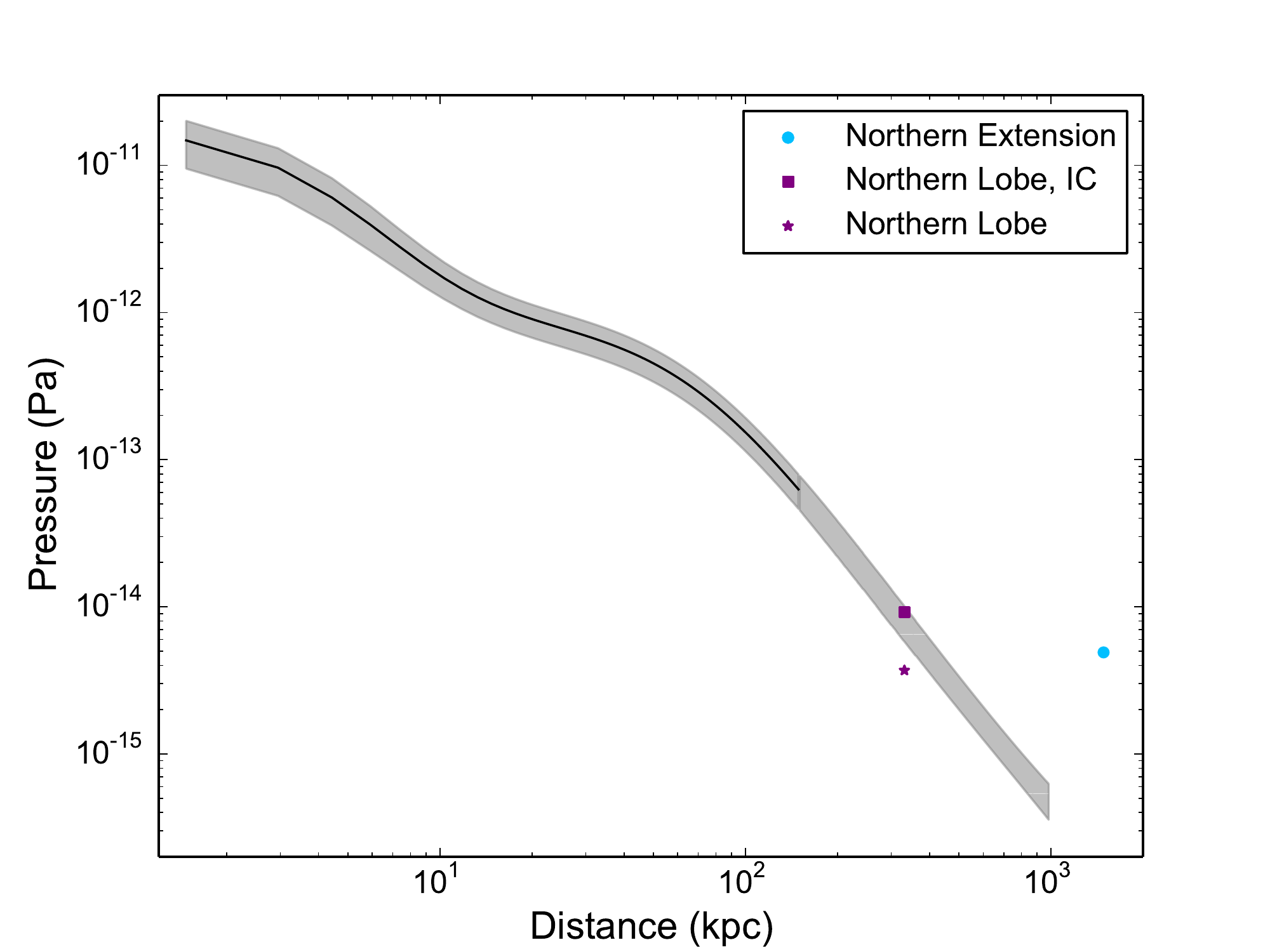}} & \subfloat[][]{\label{fig:pressure_south}}{\includegraphics[width=0.5\textwidth]{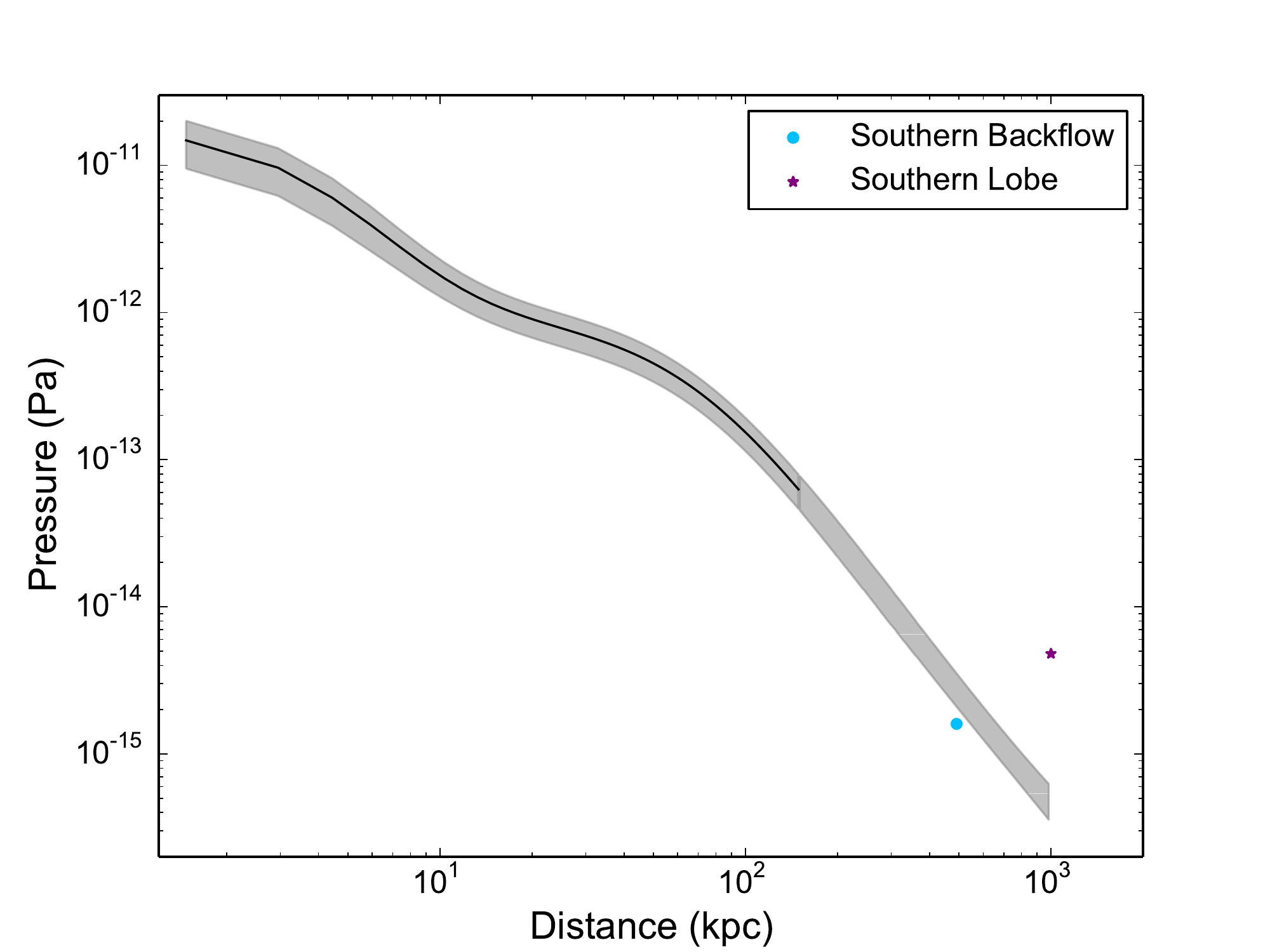}} \\
 \end{tabular}
     \caption{Internal and external pressures for NGC~6251 over a range of distances from its core.  The solid black line shows the external pressure calculated from the thermal X-ray emission, with uncertainty shaded grey, and the hatched region showing an extrapolation where there is no direct observation of the environment.  (a) Internal pressure of the northern lobe and corresponding extension, compared with a value calculated from inverse-Compton (IC) measurements \citep{Takeuchi2012}.  (b) Internal pressure of the southern lobe and southern backflow.}
     \label{fig:pressure_plot}
\end{figure*}

In Fig.~\ref{fig:pressure_plot} we compare the calculated internal pressure with the external pressure as measured from thermal X-ray observations \citep{Evans2005}.  As the volumes we calculate are highly uncertain for components containing unresolved emission, we restrict this figure, and our discussion below, to the lobes and their corresponding extension/backflow, for which the volumes are better defined.  The external pressures derived from X-ray observations are constrained only out to 150\,kpc, beyond which we extrapolate using a 2-component $\beta$ model \citep{Croston2008}.  We find that, assuming equipartition, the northern lobe is underpressured by a factor $\sim 2$ and its extension overpressured by a factor $\sim 10$, while the southern lobe is overpressured by a factor $\sim 10$ and its backflow underpressured by a factor $\sim 1.7$.  We note that the internal, equipartition pressure of the western lobe reported by \citet{Evans2005} and reproduced by \citet{Croston2008} appears to be too low, due to an incorrect low-frequency flux measurement used by those authors.

If projection effects are taken into account, both northern and southern lobes may be further from the group centre than their projected distances, placing them in a lower-pressure environment and thus giving them higher pressures relative to their surroundings.  The northern lobe, underpressured at its projected distance of $\sim 330$\,kpc, would be at pressure balance under equipartition assumptions if its true position were $\sim 440$\,kpc from the group centre.  This would place it on the axis of a straight jet at an angle to the line of sight of 41~degrees, or more or less than this if the jet bends, which appears likely.  The southern lobe, overpressured at its projected distance of $\sim 1000$\,kpc, would be more strongly overpressured if it is projected out of the plane of the sky.

Projection effects also come into play when calculating the volume of the lobes.  \citet{Evans2005} argue that the axis of the northern lobe is close to the plane of the sky based on an observed discontinuity in the X-ray surface brightness, in which case projection effects will have a minimal impact on its volume.  The southern lobe, however, could be substantially larger than we have calculated if it is elongated along the line of sight.  To take an extreme case, if it were a cylinder with a length of 1000\,kpc aligned along the line of sight, it would have an increased volume of $7.9\times10^{66}$\,m$^{3}$ and a decreased pressure of $2\times10^{-15}$\,Pa, leaving it overpressured only by a factor of~4.

Another source of uncertainty is our assumption that $\gamma_{\rm min}=10$.  To test the dependence of our results on this value we have rerun the calculations assuming $\gamma_{\rm min}=1$ and $\gamma_{\rm min}=10^{3}$ for the northern lobe and southern lobe.  Lower values of $\gamma_{\rm min}$ imply higher values for the internal pressure, and vice versa.  The northern lobe, which is underpressured by a factor of~2 in our default case, is underpressured by a factor of 1.6--5 across this range in $\gamma_{\rm min}$.  The southern lobe, which in the default case is overpressured by a factor of~10, is overpressured by a factor of 4--12 across the same range.

We conclude that the internal conditions of the radio lobes and their relationship to the environment cannot be well constrained in the absence of deep X-ray data covering the regions around both radio lobes, and given the asymmetric and poorly-constrained large-scale source geometry; however, our results suggest that internal conditions in the outer lobes of NGC~6251 may be more similar to those of FRII radio galaxies \citep[e.g.][]{Croston2018}, without the need for a large proton contribution, contrary to the conclusions of \citet{Evans2005} and \citet{Croston2008}.  One possible scenario is that the inner jet has only recently developed an FRI-like dissipative structure, and the lobe composition (as well as the presence of `warm spots' in the lobes) indicate that the source could have been fed by an FRII-like jet until relatively recently and for much of its lifetime.

\subsubsection{Spectral age}
\label{sec:specage}

The areas with the steepest spectral indices in NGC~6251 are the extension of the northern lobe and the region between the core and the southern lobe, which we have referred to as a backflow.  This label is motivated by the implied age of the material: if it has passed through the southern lobe and is flowing back toward the group centre, this would explain why its implied age is greater than that of the lobe proper.  One could also, however, construct a model in which this region contains material directly from the southern jet which has been deposited before the formation of the current southern lobe.

As these steep-spectrum components --- the northern extension and the southern backflow --- are clearly detected only at 150\,MHz, we cannot fit for a break frequency; we instead assume that this frequency $\nu_{\rm b}$ lies somewhere below 325\,MHz.  We calculate the age $t$ of both components following \citet{Alexander1987} so that
\begin{equation}
	\frac{t}{\rm Myr} = \frac{
	 1590 \, \left( \frac{B}{\mu {\rm G}} \right)^{0.5}
	}{
	 \sqrt{ \frac{\nu_{\rm b}(1+z)}{\rm GHz} } \left(
	  \left(\frac{B}{\mu {\rm G}}\right)^2 +
	  \left(\frac{B_{\rm m}}{\mu {\rm G}}\right)^2
	 \right)
	},
\end{equation}
where $B_{\rm m} = 3.18\left(1+z\right)^{2}$ is the equivalent field strength of the cosmic microwave background radiation assuming the present day temperature of 2.726\,K.  Using the equipartition magnetic fields in Table~\ref{tab:pressure} (see Section~\ref{sec:pressure}) this places lower limits of $t \gtrsim 250$\,Myr for the age of the northern extension and $t \gtrsim 210$\,Myr for the southern backflow.  The data for both the northern lobe and southern lobe show no sign of a break.  Taking 10\,GHz as a lower limit for the break frequency we find that the ages of both the northern and southern lobes have an upper limit of $t \lesssim 40$\,Myr.

\subsubsection{Group environment}

The north-south asymmetry found for the lobes of NGC~6251 in Section~\ref{sec:pressure}, with the northern and southern lobes respectively marginally under-pressured and significantly over-pressured at equipartition, might be explained by invoking asymmetry in the group environment.  If the large-scale atmosphere is not symmetric about NGC~6251 as we have assumed, but the external pressure profile instead flattens at large radii around the southern lobe only, the southern lobe might instead be much closer to pressure balance.  It has been suggested that the asymmetries seen in radio galaxies are due to environmental effects \citep{Pirya2012,Schoenmakers2000,Lara2004}.  The southern jet in NGC~6251 terminates 2.2~times further from the core than the northern jet.  \citet{Chen2011} show that the galaxy overdensity is larger in the direction of the shorter main jet of NGC~6251.  There is therefore reason to believe the environment of NGC~6251 could be asymmetric.  However, the current available X-ray data is not sufficient to investigate this directly.

 A separate estimate of the particle energetics comes from observations of inverse-Compton emission.  \citet{Takeuchi2012} fit a model to radio, X-ray and gamma--ray data.  They find the combined data are best fitted with a magnetic field in the lobe of $B=0.37$\,$\mu$G (approximately 3~times smaller than the equipartition magnetic field calculated in Section~\ref{sec:pressure}) and an injection spectral index of $\alpha=-0.5$, which breaks to $\alpha=-0.75$ at $E_{\rm b}=1.5\times10^{9}$\,eV.  This gives an energy ratio of $U_{\rm e} / U_B=45$ and an internal pressure of $8.5\times10^{-15}$\,Pa.  This would place the lobe in pressure balance with the external environment at the projected distance of the lobe from the cluster centre.  The region used to calculate this pressure includes both the lobe and the hotspot.  Given that the electron population in the hotspot and the lobe would be expected to have different characteristics, the pressure calculated by \citeauthor{Takeuchi2012} is likely to be an overestimate, and so it seems likely that the true internal pressure may be somewhere between the equipartition pressures calculated in this paper and by \citeauthor{Takeuchi2012}.

The northern extension and southern backflow are the oldest components in NGC~6251.  As such, while it is possible that the lobes are somewhat over-pressured, the extended tail-like regions of the extension and backflow are more likely to be in equilibrium with the environment.  The host galaxy group has an estimated $r_{200}$ of 875\,kpc \citep{Croston2008}.  The northern lobe reaches a projected distance of $\sim r_{500}$ while the southern lobe reaches beyond $r_{200}$.  The extended structure of NGC~6251 is therefore probing the outskirts of the group environment and the large-scale structure beyond.  The internal pressure of the extension, if at equipartition with no significant proton contribution, implies an environmental pressure of $4.9\times 10^{-15}$\,Pa, and the internal pressure of the southern backflow implies an environmental pressure of $1.6\times10^{-15}$\,Pa.  \citet{Malarecki2015} find similar pressures for 12~GRGs, and show that this pressure corresponds to the densest 6\% of the WHIM.

\subsection{Polarisation analysis}

The detection of polarised emission at low frequencies such as the 150\,MHz observations presented in this work is an effective means to precisely reconstruct Faraday depths, which has driven a great deal of recent activity.  \citet{Mulcahy2014} presented the first detections of extragalactic polarisation with LOFAR using RM~synthesis and found approximately 1~source per 1.7\,deg$^2$.  \citet{Orru2015} also report the detection of polarisation in the outer lobes of the double-double radio galaxy B1834+620.  \citet{VanEck2018} published a catalogue of 92~polarised sources at 150\,MHz in the LOFAR Two-meter Sky Survey (LOTSS) preliminary data release region.  Polarised sources have also been detected at these frequencies with the MWA: \citet{Riseley2018} published a catalogue of 81~polarised sources in the POlarised GLEAM Survey (POGS) corresponding to $\sim$ 1~source per 79\,deg$^{2}$.

Our results in Section~\ref{sec:pol} show polarised emission in the region of the bright knot in the main jet, as well as a small region of patchy polarisation in the northern lobe.  All polarisation in the inner part of the jet is depolarised, due to RM~gradients in this region \citep{Perley1984}.  It is likely that the majority of the Faraday rotation observed is due to our Galaxy, but estimating the exact value of this contribution is difficult.  Higher-frequency, high-resolution data presented by \citeauthor{Perley1984} (their Fig.~20b) show that beyond 180\,arcsec (89\,kpc) from the core the average Faraday depth is $-48.9\pm0.2$\,rad\,m$^{-2}$, which they suggest to be the Galactic contribution.  \citet{Oppermann2015} reconstruct a map of the Galactic Faraday contribution using observations of extragalactic sources.  This reconstructed map has an average Faraday rotation of $-31.6$\,rad\,m$^{-2}$ in the region of NGC~6251.  These values suggest that the extragalactic contribution is of order 1--10\,rad\,m$^{-2}$.

The Faraday-depth values we measure in the knot are in good agreement with those found by \citet{Perley1984}.  Due to LOFAR's high resolution in Faraday space it is possible to confirm that the Faraday depth is truly flat in this region with an average of $-50.97$\,rad\,m$^{-2}$ and a standard deviation of $0.07$\,rad\,m$^{-2}$.  This corresponds to the bright knot region in the LOFAR images and is the only strong detection of polarisation in our LOFAR HBA observations of NGC~6251.

\subsubsection{Limit on the group magnetic field}
\label{sec:Blim}

The detection of Faraday-thin polarised emission in the knot in the main jet and at a point in the northern lobe constitutes two measurements of the structure function, or the variation in Faraday depth as a function of physical scale.  The detection of continuous Faraday-thin emission with a variance of $\sigma_{\rm RM}^2=5\times10^{-3}$\,rad$^{2}$\,m$^{-4}$ (see Section~\ref{sec:QUfitting}) across the knot, which has a size of 2\,arcmin or 60\,kpc in projection, is a measurement of the structure function at scales up to this value.  Similarly, the difference in Faraday depth of $\delta_{\rm RM} = 2.8$\,rad\,m$^{-2}$ between the knot and the lobe gives us a single realisation of the structure function at a scale equal to their separation, which is 8\,arcmin or 240\,kpc in projection.

The measured variance of the Faraday depth in each case constitutes an upper limit on the variance $\sigma_{\rm RM}^2$ that results from Faraday rotation in turbulent magnetic fields in the group environment, thus allowing us to place an upper limit on the strength of these magnetic fields.  If the group environment is assumed to be composed of cells each with a uniform density and magnetic field strength but with a random field orientation then the expected variance in Faraday depth is
\begin{align}
  \sigma_{\rm RM}^{2}
   &=
     \left( \frac{e^3}{8 \pi^2 \, \varepsilon_0 \, m_{\rm e}^2 \, c^3} \right)^2 \Lambda_B \int{\left(B_{\parallel} \, n_{\rm e}\right)^{2}dl}
\end{align}
where $\varepsilon_0$ is the vacuum permittivity, $c$ the speed of light, and $e$ and $m_{\rm e}$ the charge and mass of the electron, or
\begin{align}
  \sigma_{\rm RM}^{2}
   &=
     \left( 812\,{\rm rad\,m}^{-2} \right)^2 \frac{\Lambda_B}{\rm kpc} \int \left( \frac{B_\parallel}{\mu{\rm G}} \frac{n_{\rm e}}{{\rm cm}^{-3}} \right)^2 \frac{dl}{\rm kpc} 
  \label{eq:sigmrm}
\end{align}
where $B_{\rm \parallel}=B/\sqrt{3}$ is the component of the magnetic field along the line of sight and $\Lambda_B$ is the size of the cells, related to the characteristic coherence length of the magnetic field \citep{lawler1982,Govoni2010}.

Assuming $B$ in the group environment to be constant, the integral in equation~(\ref{eq:sigmrm}) reduces to $\int n_{\rm e}^2 dl$; i.e.\ we need to know how the electron density $n_{\rm e}$ varies along the line of sight.  For this purpose, we use a profile of $n_{\rm e}$ for the group derived from X-ray observations \citep[][their equation~(1) \& Table~4]{Croston2008}, taking it to be centred on the core of NGC~6251, and integrate along the line of sight from specific positions in this profile.  For the Faraday-depth variation in the knot --- assuming the jet to be at an angle of $\sim 40$~degrees to the line of sight, based on the jet/counter-jet ratio \citep{Jones2002,Perley1984} --- we find the knot to be separated from the core by 150\,kpc in the plane of the sky and an additional 200\,kpc toward us along the line of sight, and integrate from this point.  For the difference in Faraday depth between the knot and the polarisation-detected region of the lobe, we take a point midway between them, separated from the core by 250\,kpc in the plane of the sky and 330\,kpc toward us along the line of sight.  Note that these points are beyond the X-ray observations used to produce the fitted profile, which has a maximum scale radius of 150\,kpc: we are extrapolating the profile slightly outside of its fitted range, as in Fig.~\ref{fig:pressure_plot}.

Under these assumptions, from equation (\ref{eq:sigmrm}), we find that the Faraday-depth variation in the knot implies a limit $B < 0.2 \, (\Lambda_B / 10\,{\rm kpc})^{-0.5}$\,$\mu$G on the magnetic field in the group environment at scales $\Lambda_B < 60\,{\rm kpc}$.  For the difference $\delta_{\rm RM}$ in Faraday depth between the knot and the lobe, as this constitutes only a single realisation of the magnetic turbulence, we use equation (\ref{eq:sigmrm}) with $\sigma_{\rm RM}^2 = \delta_{\rm RM}^2$ but relax the resulting limit by a factor of~2, permitting a 95\%-confidence limit against a Gaussian distribution, which gives us $B < 13$\,$\mu$G at the specific coherence-length scale $\Lambda_B = 240$\,kpc.  In practice, the model described by equation~(\ref{eq:sigmrm}), with magnetic-field structure at only a single scale, will not be a full description of the group environment, and a more detailed model may allow the above limits to be violated by a factor $\sim 2$ \citep[their Section~5.7 and Fig.~16]{Laing2008b}.


These magnetic-field limits may be compared to previous calculations for NGC~315, another GRG in a similarly sparse environment to NGC~6251.  \citet{Laing2006} find residual fluctuations in the Faraday depth of NGC~315 of order 1--2\,rad\,m$^{2}$ and suggest that, for plausible assumptions for the central density and characteristic magnetic field length, the central magnetic field would have to be $B_0=0.15$\,$\mu$G.  This is comparable to the upper limit calculated here for NGC~6251 for coherence lengths $\Lambda_B \sim 20$\,kpc.  Denser group environments such as those of 3C449 and 3C31 have central magnetic field strengths of order a few $\mu$G \citep{Laing2008,Guidetti2010} but field strength is expected to decrease with radius.

\section{Conclusions}
\label{sec:conclusion}

In this paper we have presented new observations of NGC~6251 at 150\,MHz with LOFAR HBA.  The images presented here are the highest sensitivity and resolution images of NGC~6251 at these frequencies to date, and reveal both an extension to the northern tail and backflow extending from the southern lobe towards the nucleus.  The integrated low-frequency spectral index is consistent with that measured previously at higher frequencies.  The lobes of NGC~6251 appear to be either close to equipartition or slightly electron dominated, similar to FRII sources; however, the lack of well-determined external environmental pressure measurements, and the uncertain and asymmetric large-scale geometry, mean that this conclusion is tentative.  We place lower limits on the ages of the low-surface-brightness extension of the northern lobe and the backflow of the southern lobe, which we have detected for the first time and are only visible at these low frequencies, of $t \gtrsim 250$\,Myr and $t \gtrsim 210$\,Myr respectively.  The possibility of FRII-like lobe composition together with the presence of `warm spots' in the radio lobes hint that the source could have been fed by an FRII-like jet for most of its lifetime, with the jet developing an FRI-like (dissipative) structure more recently (presumably due to a decrease in jet power).

We have presented the first detection of polarisation at 150\,MHz in NGC~6251, comprising a region of strong polarised emission in a knot in the northern jet and a weaker detection of polarisation in the diffuse emission of the northern lobe.  From these, taking advantage of the high Faraday resolution of LOFAR, we have placed upper limits on the strength of a magnetic field in the group environment with a single coherence scale $\Lambda_B$, with $B < 0.2 \, (\Lambda_B / 10\,{\rm kpc})^{-0.5}$\,$\mu$G for $\Lambda_B < 60\,{\rm kpc}$ and $B < 13$\,$\mu$G for $\Lambda_B = 240$\,kpc.

\section*{Acknowledgements}

JHC acknowledges support from the Science and Technology Facilities Council (STFC) under grants ST/M001326/1 and ST/R00109X/1.  We would like to thank Karl-Heinz Mack for providing fits images for the previously published WSRT and Effelsberg maps.  AMMS, JDB \& TMC gratefully acknowledge support from the European Research Council under grant ERC-2012-StG-307215 LODESTONE.  RM gratefully acknowledges support from the European Research Council under the European Union's Seventh Framework Programme (FP/2007--2013) /ERC Advanced Grant RADIOLIFE-320745.  PNB is grateful for support from the UK STFC via grant ST/M001229/1.  MJH acknowledges support from the UK Science and Technology Facilities Council [ST/M001008/1].  This research made use of Astropy,\footnote{http://www.astropy.org} a community-developed core Python package for Astronomy \citep{astropy:2013,astropy:2018}.  Finally, we would like to thank R.A.~Laing for extensive and helpful comments on the structure and content of this manuscript.


\bibliographystyle{mnras}
\bibliography{NGC6251}

\bsp 
\label{lastpage}
\end{document}